 \documentclass[11pt,twocolumn]{article}

 \pdfoutput=1

\usepackage[round]{natbib}   

\usepackage[english]{babel}

\usepackage{graphicx}
\usepackage{amsmath}

\usepackage{enumitem}

\usepackage{color}

\usepackage{float}
\usepackage{ifthen}

\usepackage{hyperref}

\usepackage{titlesec}

\titlespacing*{\section}{0pt}{1.1\baselineskip}{\baselineskip}

\begin{document}

\def\hmath#1{\text{\scalebox{1.5}{$#1$}}}
\def\lmath#1{\text{\scalebox{1.4}{$#1$}}}
\def\mmath#1{\text{\scalebox{1.2}{$#1$}}}
\def\smath#1{\text{\scalebox{.8}{$#1$}}}

\def\hfrac#1#2{\hmath{\frac{#1}{#2}}}
\def\lfrac#1#2{\lmath{\frac{#1}{#2}}}
\def\mfrac#1#2{\mmath{\frac{#1}{#2}}}
\def\sfrac#1#2{\smath{\frac{#1}{#2}}}

\def\pow{^\mmath}



\twocolumn[

\begin{center}
{\bf \Large {
Definition of the moist-air exergy norm:}}
\vspace{2mm}
\\
{\bf \Large {
a comparison with existing ``moist energy norms''.}}
\\
\vspace*{3mm}
{\Large by Pascal Marquet$\:{}^{(1)}$, 
        Jean-Fran\c{c}ois Mahfouf$\:{}^{(1)}$,
        Daniel Holdaway$\:{}^{(2)}$}. \\
\vspace*{3mm}
{\large ${}^{(1)}$ M\'et\'eo-France CNRM/GMAP / CNRS UMR3589.
 Toulouse. France.}
 \\
{\large ${}^{(2)}$ Global Modeling and Assimilation Office, 
NASA Goddard Space Flight Center, Greenbelt, 
Maryland and University Corporation for Atmospheric 
Research, Boulder, Colorado.
}
\\ \vspace*{2mm}
{\large  \it E-mail: pascal.marquet@meteo.fr}
\\ \vspace*{2mm}
{\large \it 
Submitted first to the {\bf Monthly Weather Review} 
in March 21, 2019. Revised in September 12
}
\\ \vspace*{2mm}
{\large \it
and November 20, 2019; accepted November 22, 2019.
{\bf doi:10.1175/MWR-D-19-0081.1}
}
\vspace*{1mm}
\end{center}




\begin{center}
{\large \bf Abstract}
\end{center}
\vspace*{0mm}

\hspace*{7mm}
This study presents a new formulation for the norms and scalar products used in tangent linear or adjoint models to determine forecast errors and sensitivity to observations and to calculate singular vectors.
The new norm is derived from the concept of moist-air available enthalpy, which is one of the availability functions referred to as exergy in general thermodynamics.
It is shown that the sum of the kinetic energy and the moist-air available enthalpy can be used to define a new moist-air squared norm which is quadratic in: 1) wind components; 2) temperature; 3) surface pressure; and 4) water vapor content.
Preliminary numerical applications are performed to show that the new weighting factors for temperature and water vapor are significantly different from those used in observation impact studies, and are in better agreement with observed analysis increments.
These numerical applications confirm that the weighting factors for water vapor and temperature exhibit a large increase with height (by several orders of magnitude) and a minimum in the middle troposphere, respectively.

\vspace*{7mm}

%
] 

\section{\underline{Introduction}.} 
\label{section_intro}

Several inner-products, based on ``energy'' squared norms, have been used in four-dimensional variational assimilation tools to minimize cost functions \citep{Talagrand_1981,Courtier_87,Thepaut_Courtier_1991}.
It was supposed that the ``energy'' corresponding to observational errors could be distributed equally among these different basic prognostic fields. 
Inner-products based on these ``energy'' squared norms are used to define dry semi-implicit operators and dry normal modes of GCMs or NWP models, as long as they are invariant by the linear set of primitive equations \citep{Thepaut_Courtier_1991}.

Here, the term ``energy''  means that the sum of quadratic terms is considered for perturbations of the wind components $(u')^2+(v')^2$,  temperature $(T')^2$ and surface pressure $(p'_s)^2$ or $[ \, \ln(p_s)' \, ]^2$ (see appendix~A for the list of symbols).
Moist-air generalizations of the ``energy'' squared norm have been suggested by \citet[hereafter C87]{Courtier_87}, \citet[hereafter E99]{Ehrendorfer_al_1999} or \citet[hereafter MB07]{Mahfouf_Bilodeau_2007}, among others, by including the water vapor content via an additional quadratic term $(q'_v)^2$.

The same inner-products and norms are currently used for computing dry or moist singular vectors and for determining forecast errors or sensitivity to observations based on tangent linear and adjoint models \citep{Buizza_Palmer_1995,Palmer_al_98,
Mahfouf_Bilodeau_2007,Janiskova_Cardinali_2017}.

However, all these norms suffer from a lack of consistency with physical relationships in thermodynamics, because:
i)
these ``energy'' squared norms are not based on the standard definition of energy as expressed in general thermodynamics;
ii) 
the use of the squared norm for water including the quadratic term $(q'_v)^2$ is poorly justified;
iii)
these definitions are not unique, with for instance an arbitrary tuning parameters which is often left undetermined for the water component.

Ideally, all these quadratic terms should be derived from some general laws of physics.
This is true for the average of the kinetic energy $(\overline{u^2}+\overline{v^2})/2$, 
which is the sum of the terms $(\overline{u})^2/2+(\overline{v})^2/2$ computed with the mean state of wind plus the terms $\overline{(u')^2}/2+\overline{(v')^2}/2$ computed with the perturbations of wind.
This result is true if $u'=u-\overline{u}$ and $v'=v-\overline{v}$, leading to $\overline{(u')}=\overline{(v')}=0$.
The squared norm for the wind components is computed in assimilation, singular vector and sensitivity studies with $\overline{(u')^2}/2+\overline{(v')^2}/2$, where for instance $u'$ and $v'$ are the unbiased  departures between analyses and short-range forecasts.

In contrast, the usual temperature component of the squared norm 
$\overline{\left(T^2\right)}/2 = 
\left(\overline{T}\right)^2 \!\! / \, 2 
\: + \: 
\overline{(T')^2}/ \, 2$ 
cannot be derived from the general definition of the energy and  the first law of thermodynamics.
Indeed, the dry-air internal energy or enthalpy varies linearly with temperature, with $h \approx c_{pd} \: T$ for the enthalpy up to constant reference values.
Consequently, the true energy and enthalpy cannot generate quadratic terms, due to 
$\overline{h'} = c_{pd} \: \overline{T'} = 0$.

In order to derive quadratic squared norms in both wind components and temperature, a relevant method might be based on the study of the sum of the kinetic energy and ``a form of the Available Potential Energy'' (APE) of \citet{Lorenz_1955}.
This method is chosen in \citet{Talagrand_1981}, the old ARPEGE-IFS documentation (1989, unpublished), \citet{Joly_Thorpe_1991}, \citet{Joly_1995}, \citet{Ehrendorfer_Errico_1995}, \citet{Errico_Ehrendorfer_95}, E99, \citet{Ehrendorfer_2000}, \citet{Errico_2000} and \citet{Descamps_al_2007}.

In these studies, the specific value of the approximate APE is written as $\overline{(T')^2}/(2 \: \overline{\Gamma})$, where both the perturbation of temperature $T' = T - \overline{T}$ and the stability parameter $\overline{\Gamma}$ depend on $\overline{T}$, where
\vspace{-2mm}
\begin{align}
  \overline{\Gamma}  \; &  = \; 
              \frac{\overline{T}}{c_{pd}} 
              \; - \;
              \frac{p}{R_d} \: 
              \frac{\partial \, \overline{T}}{\partial \, p} \; 
          \label{def_sigma}
\: .
\end{align}
The calculations of $\overline{\Gamma}$ are explicitly performed in \citet{Talagrand_1981} and \citet{Descamps_al_2007} by using a standard atmosphere for defining a reference profile $\overline{T}(p)$ which varies with height.

On the other hand, the stability parameter is often computed by using a constant reference value for $\overline{T}$, which is denoted by $T_r$ or an equivalent.
This leads to 
$\partial \, \overline{T} / \, \partial \,p=0$ in (\ref{def_sigma}) and to $ \overline{\Gamma} = T_r/{c_{pd}}$.
This is an explanation for the quadratic term 
\vspace{-2mm}
\begin{align}
\frac{\overline{(T')^2}}{2 \: \overline{\Gamma}}
& = \:
c_{pd} \: \frac{\overline{(T')^2}}{2 \: T_r }
\label{eq_tempe_norm}
\end{align}
which is retained in almost all present formulations of the temperature component of norms. 
A constant value $T_r$ is used in 
\citet{Courtier_87},
\citet{Thepaut_Courtier_1991},
\citet{Buizza_al_93},
\citet{Ehrendorfer_Errico_1995},
\citet{Buizza_al_96},
\citet{Mahfouf_Buizza_96},
E99, 
\citet{Errico_2000},
\citet{Barkmeijer_al01},
\citet{Zadra_al_2004},
\citet{Errico_al_2004},
\citet{Mahfouf_Bilodeau_2007},
\citet{Riviere_al_2009},
\citet{Holdaway_al_2014},
\citet{Janiskova_Cardinali_2017},
among others.

However, it is worth noting that 
the use of a constant value $T_r$ for $\overline{T}(p)$ in (\ref{eq_tempe_norm}) is not compatible with the stability term (\ref{def_sigma}) that appears in the formulation of APE expressed with pressure coordinate, where $\overline{T}$ must be defined as the isobaric average of $T$ according to \citet{Lorenz_1955}.
No other definition is allowed, and the use of a constant temperature $T_r$ makes the theory incompatible with that of Lorenz and weakens the theoretical basis for present formulations of the norm for temperature.

All temperature, pressure and water vapor components of existing squared norms correspond to the quadratic terms $(T')^2$, $(p'_s)^2$ or $[ \, \ln(p_s)' \, ]^2$ and $(q'_v)^2$.
It is thus tempting to consider these components as forming a ``total  energy'' squared norm.
However, it is explained in \citet{Errico_2000} that these squared norms are not based on clear thermodynamic definitions nor on any obvious energy norm of pressure or moisture
(``{\it Although it is called a measure of the energy, it has not been demonstrated that it is indeed such in the contexts to which it has been applied. 
The fact that it has units of energy per unit mass does not by itself qualify it as a measure of energy\/}'').
Moreover, the moist-air generalization of the APE by \citet{Lorenz_1978,Lorenz_1979} does not lead to any easy-to-use analytical formulation which could replace $\overline{(T')^2}/(2 \: \overline{\Gamma})$ with a moist-air version for $\overline{\Gamma}$. 
This means that the APE approach can not be easily generalized to moist air.

Therefore, other ideas had to be tested in order to solve the problems described so far.
Since the temperature component (\ref{eq_tempe_norm}) is presently derived from an approximate version of the APE of Lorenz, which was improved by \citet{Pearce_1978} and \citet{Marquet91} for the dry air, and then by \citet[hereafter M93]{Marquet93} for the moist air, this article examines the possibility of deriving the quadratic terms in temperature, pressure and water content from a general principle based on the concept of ``moist available enthalpy'' defined in M93.

The available enthalpy is one form of what is known as ``exergy'' in general thermodynamics.
This new exergy norm is used in \citet{Borderies_2019} to measure the relative impact of the assimilation of observations on the analysis and short-term forecasts for the French AROME model, with a large impact of the new water-content quadratic term.
Indeed, the weighting factors of the exergy norm are significantly different from those used up to now in dry and moist squared norms, in particular by several orders of magnitude for the water content.

In order to achieve some numerical validation of the theoretical formulations for the exergy norm, the same comparisons of the squared norms with inverse analysis increment estimates are made as in MB07.

The motivations for these comparisons can be found in \citet{Errico_al_2004}, where a moist norm was used with weights ``proportional to estimates of the variances of analysis uncertainty''.
It was also explained in \citet{Barkmeijer_al01} that ``in the case of forecast-error covariance prediction, a norm at initial time based on the analysis-error covariance matrix is the more appropriate'' \citep{Ehrendorfer_Tribbia_97,Palmer_al_98,Barkmeijer_al98}.
At that time, ``the analysis-error covariance metric became the reciprocal of the total-energy metric currently used at ECMWF to compute singular vectors for the EPS'' \citep{Barkmeijer_al98}.
And ``a specific-humidity norm based on error variances'' was experimented by \citet{Derber_Bouttier_99} at ECMWF, leading to a specific-humidity norm defined in \citet{Barkmeijer_al01} from the ECMWF ``averaged error variances for $q_v$'', with a strong decrease of this norm above $500$~hPa, a property that has remained unexplained until now.

This paper is organized as follows.
Existing moist-air squared norms are recalled in section~\ref{subsection_theory_existing_norms}. 
Section~\ref{subsection_exergy} presents some theoretical motivations for the use of exergy functions based on the concepts of relative entropy and Kullback-Leibler divergence.
The derivations of the moist-air available-enthalpy are conducted in Appendix~B to~G and the corresponding quadratic approximate squared norm components are shown in section~\ref{subsection_theory_new_norm} for temperature, pressure and water.
The datasets from the Canadian Meteorological Centre (CMC), the NASA Goddard Earth Observing System (GEOS) and the French ARPEGE models are described in section~\ref{section_data_method}.
These datasets are used to compare the norm components for water and temperature with the Root Mean Square (RMS) of analysis increments, with cross-sections and vertical profiles shown in section~\ref{subsection_result_ARPEGE_water} to \ref{subsection_result_CMC_GEOS} for the three models,
leading to an explanation of the decrease with height of the water vapor exergy terms described in section~\ref{subsection_result_decrease_wq}.
Forecast observation impacts are described in section~\ref{subsection_result_FSOI} for the GEOS model.
Conclusions are drawn in section~\ref{section_conclusion}.

 \section{\underline{Theoretical considerations}.} 
 \label{section_Theory}

 \subsection{\underline{Existing moist-air energy norms}.} 
 \label{subsection_theory_existing_norms}

A moist squared norm is defined in E99 by
\vspace{-0.15cm}
\begin{align}
 \! \! \!
 \! \!
   N_{\rm E99} 
      & =  \int\!\!\!\!\int\!\!\!\!\int \:
              \frac{(u')^2+(v')^2}{2} \:
              \frac{dm}{\Sigma} 
         + \! \int\!\!\!\!\int \: 
              \frac{R_d \, T_r}{g \: p_r} \: 
              \frac{{(p'_s)}^2}{2} \:
              \frac{d\Sigma}{\Sigma}
    \nonumber \\
      & \;\;\;
         + \!  \int\!\!\!\!\int\!\!\!\!\int \:
              \frac{c_{pd}}{T_r} \: 
              \frac{(T')^2}{2} \:
              \frac{dm}{\Sigma}
    \nonumber \\
      & \;\;\;
         + \! \int\!\!\!\!\int\!\!\!\!\int
              \frac{w_q(z) \: {(L_v)}^2 }{c_{pd}\:T_r} \:
              \frac{{(q'_v)}^2}{2} \: 
              \frac{dm}{\Sigma} \: . 
    \label{def_N99}
\end{align}
The state vector is represented by the local departure from mean values of basic quantities, denoted by $u'$, $v'$, $T'$, $p'_s$ and $q'_v$.
The differential mass $dm = \rho \: d\tau$ is equal to $dp\:d\Sigma/g$, where $\Sigma$ is the horizontal surface area.
The volume integrals over $dm/\Sigma$ and the surface integral over $d\Sigma/\Sigma$ represent energies per unit of horizontal area, all expressed in units of J~m${}^{-2}$.
The pressure component is expressed in E99 as a volume integral of $R_d \: T_r \: {(p'_s)}^2/(2 \: p_r^2)$, but the  expressions are equivalent providing that $\overline{p_s} \approx p_r$.

The surface pressure contribution of the squared norm is often expressed differently, in terms of the logarithm of surface pressure, leading to
\vspace{-0.15cm}
\begin{equation}
          \int\!\!\!\!\int
              \frac{R_d \: T_r \:p_r}{g} \:
              \frac{[\:\{\:\ln(p_s)\:\}'\:]^2}{2} \;
          \frac{d\Sigma}{\Sigma} \: .
          \label{def_lnps}
\end{equation}
This formalism is retained in C87, \citet{Thepaut_Courtier_1991}, \citet{Buizza_al_93}, Buizza and Palmer (1995),  \citet{Rabier_al_1996}, \citet{Palmer_al_98}, \citet{Errico_2000}.

The two formalisms using the surface pressure or its logarithm are nearly equivalent, providing that $\overline{p_s} \approx p_r$.
Indeed, the departure term must be computed as $ \{\:\ln(p_s)\:\}' = \ln(p_s)-\overline{\ln(p_s)}$ in (\ref{def_lnps}) and the perturbation of pressure is equal to $p'_s \:=\: p_s-\overline{p_s}$ in (\ref{def_N99}), leading to 
$ \{\:\ln(p_s)\:\}' = \ln(1+p'_s / \overline{p_s}) - \overline{ \ln(1+p'_s / \overline{p_s}) } \: \approx \: p'_s /\overline{p_s}$ up to small higher order terms.

The justification for the last integral of (\ref{def_N99}) depending on the variance of water vapor content can be found in \citet{Ehrendorfer_al_95}, \citet{Buizza_al_96}, \citet{Mahfouf_al_96} and E99.
The water contribution of the squared norm is derived from the temperature contribution $c_{pd}\:(T')^2/(2\:T_r)$ with the additional hypothesis that changes of temperature and moisture are related by $c_{pd}\:T' \approx - L_v\:q'_v$, namely by assuming a conservation of the moist static energy 
$c_{pd}\:T + L_v\:q_v + \phi$
at constant height for all moist (condensation) process.
A similar quadratic term was suggested in C87, where two scale factors for height ($H_r$) and water content $(Q_r)$ were defined, leading to the equivalent formulation $\, g \: H_r \: (q'_v)^2  \, / \, (Q_r)^2 \,$.

The question addressed in E99 is the relevance of that special formulation for the water contribution.
Due to the uncertainty in the assumption $c_{pd}\:T' + L_v\:q'_v \approx 0$ (particularly in frequently under-saturated moist areas without condensation processes), 
an additional relative weight $w_q(z)$ (also denoted by $w^2$ or $\epsilon$, depending on papers) is added in the last integral of (\ref{def_N99}).
The effects of making this relative weight larger or smaller than the standard value $1$ are discussed in E99 and \citet{Barkmeijer_al01}, where $w_q(z)$ may increase with height in the upper troposphere and in the stratosphere.

An alternative definition of the water contribution of the squared norm is proposed in MB07 by replacing the assumption of conservation of perturbed moist static energy by a conservation of relative humidity approximated by $q_v/q_{sw}$.
This assumption is expected to be realistic in cloudy areas where relative humidity reaches $100$~\%, however it may not be realistic in frequently under-saturated moist areas.
The constraint of zero departure (at constant pressure) in the quantity $q_v / q_{sw}(T,p)$ corresponds to  $q'_v = \: ( \overline{\Gamma}_q ) \: T'$, where 
$\overline{\Gamma}_q = \overline{q_v} \; 
\partial \ln(\overline{q_{sw}}) / \partial \, T$.
The alternative contribution proposed in MB07 can be written as 
\vspace{-0.15cm}
\begin{equation}
          \int\!\!\!\!\int\!\!\!\!\int
       \: \frac{c_{pd}}{T_r}
       \: \frac{1}{(\overline{\Gamma}_q)^2}
       \: \frac{{(q'_v)}^2}{2}
       \: \frac{dm}{\Sigma} \: . 
          \label{def_Nqv_MB07}
\end{equation}
MB07 found that this revised formulation for the water component of the norm better match the RMS of the analysis increments than the E99 norm.
Indeed, the MB07 formulation better reflects the typical size of perturbations produced by data assimilation systems and (\ref{def_Nqv_MB07}) accounts for the exponential decrease of specific humidity with altitude, leading to much smaller absolute errors than with the original constant contribution in the last integral of (\ref{def_N99}).
This result agrees with the increase of $w_q(z)$ with altitude considered in \citet{Zadra_al_2004} in moist singular vector computations.
The aim was to suppress the impact of humidity perturbations in the stratosphere according to the results of \citet{Buizza_al_96} and E99, who showed that for increasing $w_q$ the contribution of the dry fields dominates initially, whereas the contribution of moisture dominates at the final time (and vice versa when $w_q$ is smaller).

According to \citet{Errico_al_2004} and MB07, the grid-point discretization of either  (\ref{def_N99}) or (\ref{def_Nqv_MB07}) can be written as the inverse variance weighted squared norm 
\vspace{-0.15cm}
\begin{align} 
\!\!\!\!\!
  & \sum_{ijk}
    \left( 
       \frac{(u'_{ijk})^2}{V_u}
     +
       \frac{(v'_{ijk})^2}{V_v}
     +
       \frac{(T'_{ijk})^2}{(V_{T1})_{jk}}
     \right)
    \omega_{ij} \; \Delta \sigma_k
    \; + \:
 \nonumber \\
 \!\!\!\!\!
  &
    \sum_{ij}
    \left(
       \frac{(p_s')_{ijk}^2}{(V_{p1})_{jk}}
    \right)
    \omega_{ij}
     \: + \:
     \sum_{ijk}
    \left(
       \frac{(q_v')_{ijk}^2}{(V_{q1})_{jk}}
    \right)
    \omega_{ij} \; \Delta \sigma_k
    \label{def_N99_V} \: ,
\end{align}
where $\Delta \sigma_k$ is the thickness of the layer $k$ in the $\sigma$ vertical coordinate and $\omega_{ij}$ is the fractional coverage of the model grid box defined by the zonal ($i$) and meridional ($j$) indices.

The weighting factors $V_u$, $V_v$, $(V_{T1})_{jk}$, $(V_{p1})_{j}$ and $(V_{q1})_{jk}$ will hereafter be referred to as ``$V$-terms''. 
They are interpreted as variances of analysis errors in \citet{Errico_al_2004} and MB07.
The indices $j$ and $k$ mean that temperature, surface pressure and water variances can a priori depend on latitude ($j$) and/or altitude ($k$).

From (\ref{def_N99}) and (\ref{def_N99_V}), the $V$-terms in E99 can be written as
\vspace{-0.15cm}
\begin{align} 
\!\!\!\!\!\!\!
  & V_u = V_v = 2 = \!V_0 \: , \; \; 
    V_{T1} = V_0 \; \frac{T_r}{c_{pd}} 
    = V_0 \; \frac{(T_r)^2}{c_{pd} \: T_r} 
    \: , \label{def_N99_def_VuTp} \\
\!\!\!\!\!\!\!
  &
  V_{p1} = \!V_0 \; \frac{(p_r)^2}{R_d \: T_r} \, ,
 (V_{q1})_{k} =  
        \frac{V_0}{w_q(z)} \;
        \frac{c_{pd} \: T_r}{(L_v \: Q_r)^2} \; (Q_r)^2
      .\! \label{def_N99_def_Vq}\!\!
\end{align}
The four terms $V_u$, $V_v$, $V_{T1}$ and $V_{p1}$ are all constant, whereas $(V_{q1})_{k}$ may depend on altitude for water, via the arbitrary weight $w_q(z)$.

All terms in parentheses in (\ref{def_N99_V}) are dimensionless in \citet{Errico_al_2004} and MB07, where the dimensions of the square root of $(V_{T1})_{jk}$, $(V_{p1})_{j}$ and $(V_{q1})_{jk}$ are K, hPa and kg~kg${}^{-1}$, respectively.
The square-root of these $V$-terms will be called ``$SqV$-terms'' hereafter.
The dimensionless characteristic of (\ref{def_N99_V}) can be explained by first multiplying all terms of (\ref{def_N99}) by the dimensionless value $2$, and then by dividing all terms by the same energy term $V_0 = 2$~J~kg${}^{-1}$.
Therefore, the dimensions of $c_{pd} \: T_r$, $R_d \: T_r$ and $L_v \: Q_r$ are same as the one of $V_u = V_v = V_0$, namely in units of m${}^{2}$~s${}^{-2}$ or J~kg${}^{-1}$.
The value of the dummy specific content $Q_r$ has no impact in (\ref{def_N99_def_Vq}); it is introduced to highlight the relevant dimension of kg${}^{2}$~kg${}^{-2}$ for $(V_{q1})_{jk}$.

The definition (\ref{def_Nqv_MB07}) proposed by MB07 corresponds to
\vspace{-0.15cm}
\begin{align} 
 (V_{q2})_{jk} & = V_0 \;
          \frac{T_r}{c_{pd} \: (\overline{T})^2} \;
     \left(
            \frac{\overline{T}}{\, \overline{q_{sw}} \,}
     \:
            \frac{\partial \,\overline{q_{sw}}}{\partial \, T}
     \right)^2 \:
     \left(
             \overline{q_{v}}
     \right)^2\; ,
    \label{def_N99_def_Vq2} \\
 (V_{q2})_{jk} & \approx V_0 \;
          \frac{T_r}{c_{pd}} \:
     \left(
            \frac{L_v \: \overline{q_v}}
            {R_v \: (\overline{T})^2}
     \right)^2
     \; .
    \label{def_N99_def_Vq3}
\end{align}
From (\ref{def_N99_def_Vq2}), $(V_{q2})_{jk}$ is expressed in kg${}^{2}$~kg${}^{-2}$, because  
$c_{pd} \: (\overline{T})^2 / T_r$ has the same dimension as $V_0$.
This means that the dimension of the square root of $(V_{q2})_{jk}$ is the same as the specific content $\overline{q_v}$, which is expressed in kg~kg${}^{-1}$ and, from (\ref{def_N99_def_Vq3}), varies with altitude via the ratio of the average terms $\overline{q_v}$ and $(\overline{T})^2$.

 \subsection{\underline{Relative entropy, Exergy and} \\
 \underline{Available enthalpy}.} 
 \label{subsection_exergy}

Due to the uncertainty and plurality in $V_{T1}$, $V_{q1}$ or $V_{q2}$ defined in E99 or MB07, and due to the arbitrary values for $w_q(z)$, it is necessary to find a more general and comprehensive ``measure,'' ``norm'' or ``distance'' between a perturbed thermodynamic state defined by $(T_2, q_{v2}, p_{s2})$  and a reference one defined by $(T_1, q_{v1}, p_{s1})$.

It is explained in section~3 of \citet{Marquet_Thibaut_2018} that this distance can be measured by the quantity referred to as ``relative entropy'' by \citet{Shannon_1948} and then defined in \citet{Kullback_Leibler_1951} and \citet{Kullback_1959} by
\begin{equation}
 K( x||y  )  = 
 \sum_{j=1}^n \: x_j \:  
 \log( x_j/y_j ) \:
 \label{eq_K}
\end{equation}
where the $x_j$'s represent a real state ($x$) and the $y_j$'s a reference state ($y$) of the system \citep[see][]{Cover_Thomas_91}.\footnote{The original notations of Shannon and Kullback using $p(p_j)$ and $q(q_j)$ are replaced here to avoid confusion with the pressure $p$ and the specific water content quantities $q_t, q_v, q_l, q_i$.}

This Kullback-Leibler divergence $K$ is usually interpreted as being a non-symmetric measure of how much the $x_j$'s deviate from the $y_j$'s.
It also represents the ``gain in information'' of the state characterized by the distribution ($x_j$)  with respect to the equilibrium distribution (${y_j}$).
Therefore, it is unclear whether $K$ corresponds to the measure or the distance between the two thermodynamic states $(T_2, q_{v2}, p_{s2})$ and $(T_1, q_{v1}, p_{s1})$.

The main difficulty lies in determining the $x_j$'s and the $y_j$'s that correspond to these two thermodynamic states.
Moreover, the relative entropy $K$ is clearly different from the entropy 
$s(x) = 
- \sum_{j=1}^n \: x_j 
\: \log(x_j)$
of \citet{Shannon_1948}, with a change of sign and another reference state $y_j$ included in (\ref{eq_K}).
However, it is possible to show that the macroscopic value of $K$ roughly corresponds to the free energy function $e_i - T_r \: s$, which is different from the entropy $s$ because it  depends on the internal energy $e_i$ and a reference temperature $T_r$.
More precisely, it is shown for instance in \citet{Procaccia_Levine_1976}, \citet{Eriksson_Lindgren_1987} and \citet{Karlsson_90} that the exergy of moist air can be computed by the ``available energy'' function $a_e = k_B \: T_r \: K$, with $K(x||y)$ given by (\ref{eq_K}).
This function $a_e$ can be written in terms of the local atmospheric variables ($p$, $T$, $q_n$), leading to  
\vspace*{-1.5mm}
\begin{align}
a_e & \: = \; 
 (e_i - e_{ir}) + p_r \: (\alpha - \alpha_r) 
- T_r \: (s - s_r) 
 \nonumber \\ & \; \; \; \; \;
- \sum_n \mu_{rn} \: (q_n-q_{rn})
\: ,
\label{eq_A_K}
\end{align}
where the subscript ``$r$'' denotes a reference state and where the sum over ``$n$'' represents the dry air, water vapor, liquid water and ice species.
The specific volume is $\alpha = 1/\rho$ and the specific contents $q_n$ are multiplied by the reference Gibbs functions $\mu_{rn} = h_{rn} - T_r \; s_{rn}$.
The quantity $a_e$ given by (\ref{eq_A_K}) is called ``{\it maximum available work from a nonflow system\/}'' by \citet[Eq.5.12]{Bejan_2016} for system at rest reaching a pressure equilibrium with the environment (the laboratory).
The last sum over $n$ in (\ref{eq_A_K}) is called ``{\it chemical exergy\/}'' by Bejan, while the other terms form the ``{\it nonflow exergy\/}.''

The sum of the terms $(e_i - e_{ir})$ and $- \: p_r \: (\alpha - \alpha_r)$ in (\ref{eq_A_K}) must be replaced by the difference in specific enthalpy $(h - h_r)$ to form the 
``{\it thermomechanical and chemical flow exergy\/}'' defined in \citet[Eq.5.25]{Bejan_2016}.
It is the sameavailable enthalpy function as that studied in \citet{Marquet91} and M93 and corresponding to (\ref{defam}), with all other terms of (\ref{eq_A_K}) remaining the same, leading to
\vspace*{-1.5mm}
\begin{align}
a_m & \: = \;  (h - h_r) 
       \: - \: T_r \: (s - s_r)
       \: - \: \sum_n \mu_{rn} \: (q_n-q_{rn})
\: .
\label{eq_Am}
\end{align}
The use of the specific enthalpy $h$ to replace the internal energy is motivated by the natural application of $h$ to the flowing moist-air atmosphere.
No hypothesis is made from this point of view, since the use of enthalpy does not impose movements that would be made ``at constant pressure''.
The change in the variable $h=e_i + p/\rho$ is simply mathematical, with no underlying physical assumptions.
One of the interests of the introduction of the enthalpy $h$ is the existence of the Bernoulli function $h+g\:z+(u^2+v^2)/2$, which is constant during stationary, adiabatic and frictionless motions, with a similar
Bernoulli's law derived in M93 for $a_m+g \:z+(u^2+v^2)/2$.

The flow exergy $a_m$ given by (\ref{eq_Am}) ensures the definition of the aforementioned general distance between a perturbed atmospheric state and a reference one.
Indeed, since the available enthalpy is the maximum work (or energy) that a system can deliver when passing from a reference state to the real state, this work is produced by transformations from different forms of energy to other forms of energy.

In particular, it is shown in M93 that a Bernoulli equation exists and that the sum $a_m(T,p,q_v,q_l,q_i) + (u^2+v^2)/2 + \phi$ is conserved along any streamline of an adiabatic frictionless and reversible steady flow of a closed parcel of moist air.
This means that the conversions between the potential energy, the kinetic energy and the temperature, pressure and water components of $a_m(T,p,q_v,q_l,q_i)$ given by (\ref{eq_Am}) can be evaluated with the weighting factors $V_T$, $V_p$ and $V_q$, ensuring relevant thermodynamic transformations of energy from one form to another.

 \subsection{\underline{The new moist-air available-enthalpy}
          \\ \underline{norm}.} 
 \label{subsection_theory_new_norm}

The three components of the squared norm based on the M93 exergy function given by (\ref{eq_Am}) are derived in the Appendices~B to G.
They can be written in terms of the square of the perturbations of temperature (\ref{def_N_T}), surface pressure (\ref{def_N_p})-(\ref{def_N_p3}) and water vapor (\ref{def_N_v}), leading to
\vspace{-0.15cm}
\begin{align}
\!\!\!\!\!\!\!\!
   N_T \;
      & = \!\! \; \int\!\!\!\!\int\!\!\!\!\int\limits^{ } 
          \left[ \:
              \frac{c_{pd} \: T_r }{(\overline{T})^2 } \:
        \: \right] 
              \frac{(T')^2}{2} \;
              \frac{dm}{\Sigma} \;
      \label{def_Nm_T}  \: , \\
\!\!\!\!\!\!\!\!
   N_p \; 
      & = \; 
              \frac{R_d \: T_r}{g \; \overline{p_s}} \: 
              \frac{ \overline{{(p'_s)}^2} }{2}
      \: =
        \int\!\!\!\!\int\!\!\!\!\int\limits^{ }
          \left[ \:
              \frac{R_d \, T_r}{(\overline{p_s})^2} \:
        \: \right] 
              \frac{{(p'_s)}^2}{2} \;
             \frac{dm}{\Sigma}
      \label{def_Nm_p}  \: , \\
\!\!\!\!\!\!\!\!
   N_v \; 
      &  = 
        \int\!\!\!\!\int\!\!\!\!\int\limits^{ }
          \left[ \:
              \frac{R_d \, T_r}
  {\left( r_0 \: \overline{r_v} \right)} \:
        \: \right] 
              \frac{{(r'_v)}^2}{2} \;
             \frac{dm}{\Sigma} \;
      \label{def_Nm_v}  \: . 
\end{align}

\begin{figure*}[h]
\centering
\includegraphics[width=0.99\linewidth]{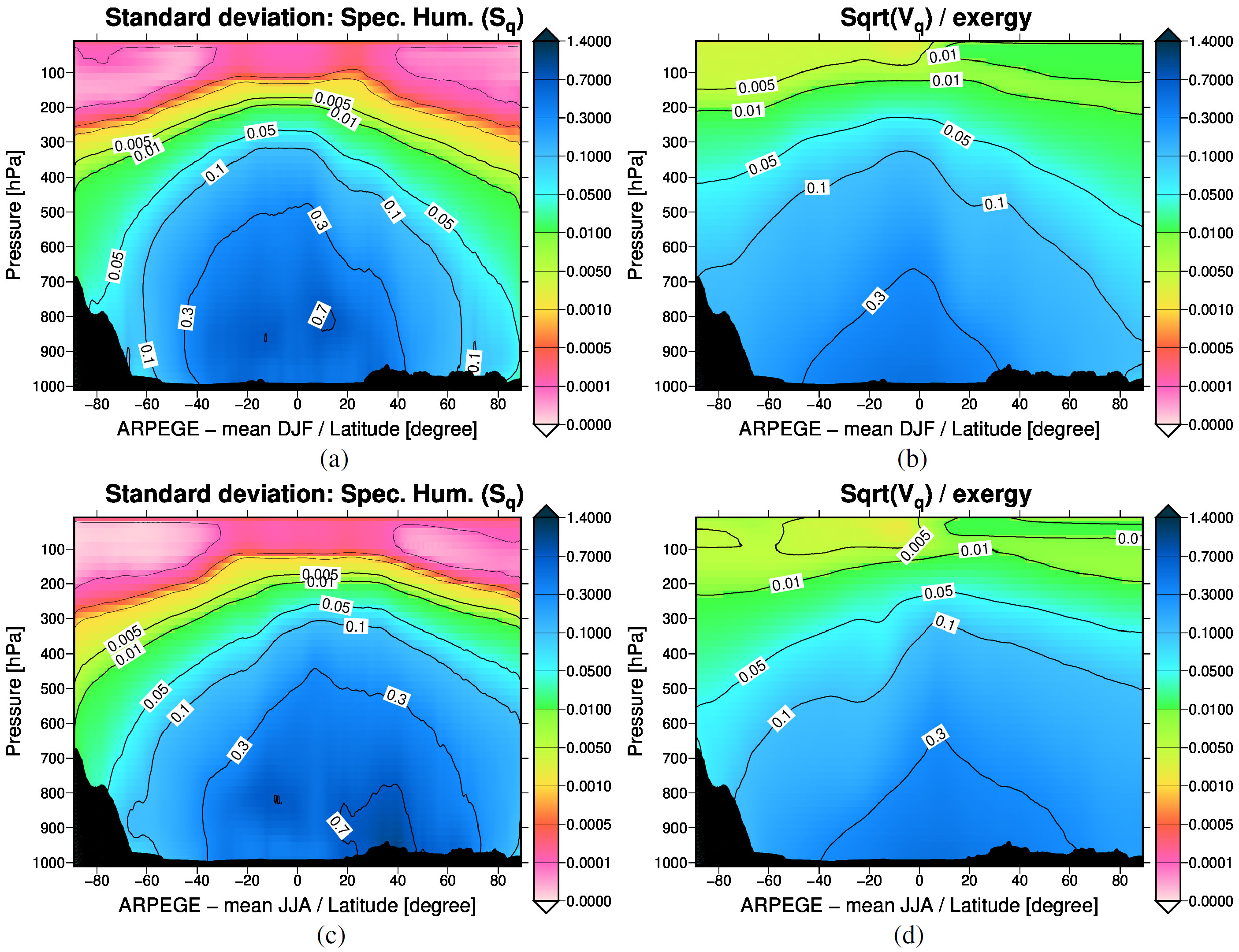}
\vspace*{-3mm}
\caption{\it \small
The seasonal averages of the RMS of analysis increments for water $S_q$ (g~kg${}^{-1}$) are computed for ARPEGE outputs every $6$~hours and plotted in latitude-pressure sections for: (a) winter (DJF); and (c) summer (JJA).
The corresponding seasonal averages of the exergy $SqV$-term $\sqrt{V_{q}}$ given by (\ref{def_Nm_def_q}) are plotted for: (b) winter (DJF); and (d) summer (JJA).
\label{fig_ARP_DJF_JJA_water_incr_anal_exe}}
\end{figure*}

The new $V$-terms corresponding to (\ref{def_N99_def_VuTp})-(\ref{def_N99_def_Vq3}) for temperature, pressure and water content can be written as
\vspace{-0.15cm}
\begin{align} 
\!\!\!\!\!\!\!\!
    (V_{T})_{jk} & = \frac{V_0 \: T_r}{c_{pd}} \:
    {\left(\frac{\overline{T}}{T_r}\right)}^{\!\!2} \: , \;
    (V_{p})_{j}    = \frac{V_0 \: p_r^2}{R_d \: T_r} \:
    {\left(\frac{\overline{p_s}}{p_r}\right)}^{\!\!2} \: ,
    \label{def_Nm_def_Tp}
   \\
\!\!\!\!\!\!\!\!
    (V_{q})_{jk} & =  
                  \frac{V_0 \:r_0 \: \overline{r_v}}{R_d \: T_r} 
              \; =
                  \frac{p_r-e_r}{e_r} \:
                  \frac{V_0 \: r_r \; \overline{r_v}}{R_d \: T_r}
              \; =
                  \frac{V_0 \: \overline{r_v}}{R_v \: T_r} 
              \: .
    \label{def_Nm_def_q}
\end{align}
From the first formulation in (\ref{def_Nm_def_q}), $(V_{q})_{jk}$ is independent of $r_r$.
The last formulation in (\ref{def_Nm_def_q}) is obtained with $R_v = R_d / r_0$ and $r_0 = r_r \: (p_r-e_r)/e_r \approx 622$~g~kg${}^{-1}$, where $r_0$ is proportional to the reference mixing ratio $r_r$.
This shows that the dimensions of $(V_{q})_{jk} $ and of $r_r \; \overline{r_v}$ are both kg${}^{2}$~kg${}^{-2}$, since $V_0=2$~m${}^{2}$~s${}^{-2}$ and $R_d \: T_r$ have the same dimension.
Therefore, the square root of $(V_{q})_{jk}$ has the dimension of a mixing ratio, as expected.

From (\ref{def_N99_def_Vq}) and (\ref{def_Nm_def_Tp}) the pressure $V$-terms $V_{p1}$ and $(V_{p})_{j}$ may be close to each other if $p_r \approx \overline{p_s} \approx 1000$~hPa, with $(V_{p})_{j}$ only depending on $\overline{p_s}$ and being independent on $p_r$.

Differently, the temperature and water $V$-terms can differ significantly because $\overline{T}$ and $\overline{r_v}$ vary with height.
This is especially true for $(V_{q})_{jk}$ since $\overline{r_v}$ may vary by $3$ orders of magnitude from the surface to the stratosphere.

The comparison of (\ref{def_Nm_def_q}) with (\ref{def_N99_def_Vq}) allows a computation of the unknown dimensionless weighting factor $w_q(z)$ in E99, leading to
\vspace{-0.15cm}
\begin{align} 
  w_q(z) & \: = \;
        \frac{c_{pd} \: R_v \: (T_r)^2}
        {(L_v)^2} 
        \:
        \frac{1}
        {\overline{r_v}(z)} 
         \, ,
    \label{def_wz} \\
  w_q(z) & \: = \;
        \frac{(c_{pd} \: T_r ) \: (R_d \: T_r)}
        {(L_v \: r_r)^2} 
        \:
        \left(
        \frac{e_r}
        {p_r-e_r} 
        \right)
        \:
        \frac{r_r}
        {\overline{r_v}(z)} 
         \, ,
    \label{def_wz2}
    \end{align}
where
$R_v = (R_d \: e_r)/ [\: (p_r-e_r) \: r_r \: ]$ is used to derive the formulation (\ref{def_wz2}), which better shows the dimensionless feature due to the compensation of the terms $c_{pd} \: T_r $ and $R_d \: T_r$ with $L_v \: r_r$, also of $e_r$ with $p_r-e_r$ and of $r_r$ with $r_v$.

The exergy weighting factor ($\ref{def_wz2}$) explains the expected behavior for $w_q(z)$, which increases with height for decreasing values of ${\overline{r_v}(z)}$.
A similar decrease holds with the MB07 value derived from the comparison of the constant relative humidity $V$-term (\ref{def_N99_def_Vq3}) with the constant MSE $V$-term (\ref{def_N99_def_Vq}), leading to
\vspace{-0.15cm}
\begin{align} 
  w_{q2}(z) & \: \approx \;
        \frac{(c_{pd} \: T_r )^{2} \: (R_d \: T_r)^{2}}
        {(L_v \: r_r)^4} 
        \:
        \left(
        \frac{e_r}
        {p_r-e_r} 
        \right)^{\! 2}
        \:
        \left(
        \frac{r_r}
        {\overline{q_v}(z)} 
        \right)^{\! 2}
         \, .
    \label{def_wz_MB07}
\end{align}
A comparison of (\ref{def_wz_MB07}) with (\ref{def_wz2}) shows that $w_{q2} \approx (w_{q})^2$ because $\overline{r_v} \approx \overline{q_v}$.
Therefore, the MB07 value is approximately the square of the available enthalpy value, leading to an enhanced variation of $w_{q2}(z)$ with height in MB07.

\begin{table*}
\caption{\it \small The reference temperatures $T_r$ (K) and pressures $p_r$ (hPa) used (from the left to the right) in:
\citet{Pearce_1978} and M93, 
\citet{Buizza_al_96} and \citet{Mahfouf_Buizza_96}, 
E99 and 
\citet{Holdaway_al_2014},
\citet{Errico_al_2004} and MB07,
\citet{Janiskova_Cardinali_2017}.
\vspace*{1mm}
\label{Table_Tr_Pr}}
\centering
\begin{tabular}{|c|c|c|c|c|c|}
\hline 
        & P78/M93     & B96/MB96  & E99/H14    & E04/MB07    &  JC17  \\ 
\hline 
 $T_r$  & $251$   & $300$ & $270$  & $300$  & $350$  \\ 
 $p_r$  & $367.8$ & $800$ & $1000$ & $1000$ & $1000$ \\
\hline 
\end{tabular}
\end{table*}

\begin{table*}
\caption{\it \small 
The reference mixing ratio $r_r (T_r, p_r)$ defined as $r_0 \; e_r(T_r) / [\: p_r-e_r(T_r) \:] $ in g~kg${}^{-1}$ and the saturated pressure $e_r(T_r)$ in hPa computed for several reference temperatures $T_r$ in K and pressures $p_r$ in hPa.
\vspace*{1mm}
\label{Table_rr}}
\centering
\begin{tabular}{|c||c|c|c||c|}
\hline 
$\!$$\!$$T_r \downarrow \setminus  \: p_r \rightarrow$$\!$$\!$ & 
$\!$$367.8$~hPa$\!$ & $\!$$800$~hPa$\!$ & $\!$$1000$~hPa$\!$ & $e_r(T_r)$ \\ 
\hline 
 $251$~K      & $r_r=1.42$  & $\!$$r_r=0.653$$\!$ & $\!$$r_r=0.522$$\!$ & $\!$$(0.838)$$\!$ \\ 
 $270$~K      & $r_r=8.11$  & $r_r=3.69$  & $r_r=2.94$  & $(4.7)$ \\
 $273.15$~K   & $r_r=10.6$  & $r_r=4.81$  & $r_r=3.84$  & $(6.11)$ \\
 $280$~K      & $r_r=17.5$  & $r_r=7.86$  & $r_r=6.26$  & $(9.9)$ \\
 $300$~K      & $r_r=70.6$  & $r_r=29.5$  & $r_r=23.3$  & $(35.3)$ \\ 
 $325$~K      & $r_r=558$   & $r_r=144$   & $r_r=107$   & $(134)$ \\ 
 $350$~K      & $--$        & $r_r=1928$  & $r_r=769$   & $(411)$ \\
\hline 
\end{tabular}
\end{table*}

Although the reference value of water content has no impact on the water term $(V_{q})_{jk}$ given by (\ref{def_Nm_def_q}), it is possible to compute, for the sake of internal consistency and realism, both $e_r$ and $r_r$ for several of the values of $T_r$ and $p_r$ which, from Table~\ref{Table_Tr_Pr}, are typically used in atmospheric research (semi-implicit algorithms, computation of singular vectors and studies of sensitivity to observations or forecast errors).
The result is shown in Table~\ref{Table_rr} for saturating pressures $e_r = e_{sw}(T_r)$ or $e_{si}(T_r)$ with respect to the more stable state (liquid water or ice), depending on the temperature $T_r$.
The zero Celsius and $280$~K temperatures are added to show the rapid increase of both $e_r$ and $r_r$ with $T_r$ for an increase of a few degrees between $270$ and $280$~K.
The higher temperature $T_r=350$~K leads to unrealistically large values of $r_r$, which are even undefined (negative) for $367.8$~hPa.
The explanation for these impossible values for some couple $(T_r, p_r)$ comes from the fact that $e_r$ is defined as the saturation pressure at the temperature $T_r$.
We therefore assume that $p_r>e_r$, which is not verified for example for $T_r=350$~K for which $e_r=411$~hPa is greater than $367.8$~hPa in Table~\ref{Table_rr}.
But this assumption $p_r>e_r$ does not limit the validity of the theory, in the same way that the assumption $p>e$ for humid air does not limit the two state equations for dry air and water vapour.
Therefore the available enthalpy function and the exergy norm are well-defined 
for values $T_r<300$~K for which the ratios $|X_v/Y_v|$ are greater than $10$ in Table~\ref{Table_Yv_over_Xv}, regardless of the pressure $p_r$.

 \section{\underline{The Datasets}.} 
 \label{section_data_method}
 
The RMS of analysis increments $S_q$ and the $SqV$-terms are computed for three systems using 3DVAR or 4DVAR algorithms.
The periods correspond to either individual days, month or seasonal periods.
The aim is to show that the temperature and water components of the exergy norm lead torobust results (i.e. that are valid for a wide range of durations and for different systems).

ARPEGE is the NWP model used at the French weather service at M\'et\'eo-France \citep{Courtier_al_1991}.
The horizontal Gauss grid is based on a Schmidt projection with a spectral  truncation T1198 and a stretching factor of $2.2$ (i.e. with a varying resolution from $7$~km over France to $33$~km over the South Pacific).
The vertical grid has 105 hybrid levels extending from $10$~m to $0.1$~hPa.
The data assimilation is based on a 6-hourly incremental 4DVAR \citep{Courtier_al_1994}, with increments computed at the truncations T149c1 ($135$~km) and T399c1 ($50$~km).

The Global Environment Multiscale (GEM) model \citep{cote_al_1998b,cote_al_1998a} studied in MB07 is used at the Canadian Meteorological Centre (CMC).
The global horizontal grid has a uniform resolution of $1.5$ degrees in longitude and latitude.
The resolution is variable in the vertical, with $28$ $\sigma$ levels extending from the surface up to $10$~hPa.
The analysis increments are diagnosed by the CMC 3DVAR system \citep{Gauthier_al_2007}.

The Goddard Earth Observing System version 5 (GEOS-5) is an atmospheric global circulation model developed by the National Aeronautics and Space Administration's (NASA) Global Modeling and Assimilation Office (GMAO). 
The model is based on the finite volume cubed-sphere (FV3) dynamical core \citep{Putman_2007}.
The Modern-Era Retrospective analysis for Research and Applications (MERRA-2) Version 2  \citep{Gelaro_al_2017} is a global reanalysis produced by GMAO using the GEOS forecast model and gridpoint statistical analysis data assimilation system \citep{Wu_al_2002,Kleist_al_2009}.
The 3D-Var system MERRA-2 produces an analysis every 6 hours from 1980 to the present day.
The horizontal resolution of the data assimilation and model is around $50$~km, or $0.5$~degree.
In the vertical, $72$ hybrid sigma-pressure levels are used, reaching from the surface to $0.01$~hPa.
The linearized version of GEOS includes the FV3 dynamical core and a linearization of the relaxed Arakawa-Schubert convection scheme \citep[hereafter H14]{Holdaway_al_2014}, single moment cloud scheme \citep{Holdaway_al_2015} and a simplified boundary layer scheme.

 \section{\underline{The results}.} 
 \label{section_Result}

 \subsection{\underline{Seasonal means of ARPEGE:} \\
             \underline{the water norms}.} 
 \label{subsection_result_ARPEGE_water}
 
The ARPEGE seasonal averages of RMS of analysis increments $S_q$ and exergy $SqV$-term $\sqrt{V_{q}}$ are shown in Fig.~\ref{fig_ARP_DJF_JJA_water_incr_anal_exe}.
The winter and summer averages are computed with data 4 times per day ($0$, $6$, $12$, $18$~UTC). 

The general patterns for $S_q$ and $\sqrt{V_{q}}$ are roughly similar, with a large vertical decrease with height (from $0.5$ to less then $0.005$~g~kg${}^{-1}$) and seasonal latitude oscillations following the regions of maximum surface temperatures (from $-15$~degree in DJF to $+15$~degree in JJA).

Values close to the ground are of the same order of magnitude for the analysis increments ($\approx 0.7$~g~kg${}^{-1}$), the exergy term ($\approx 0.4$~g~kg${}^{-1}$) and the E99 term ($\sqrt{V_{q1}} \approx 0.31 $~g~kg${}^{-1}$ or $0.57$~g~kg${}^{-1}$ computed with $T_r=300$~K and $w_q=1.0$ or $w_q=0.3$).

\begin{figure}[hbt]
\centering
\includegraphics[width=0.99\linewidth]{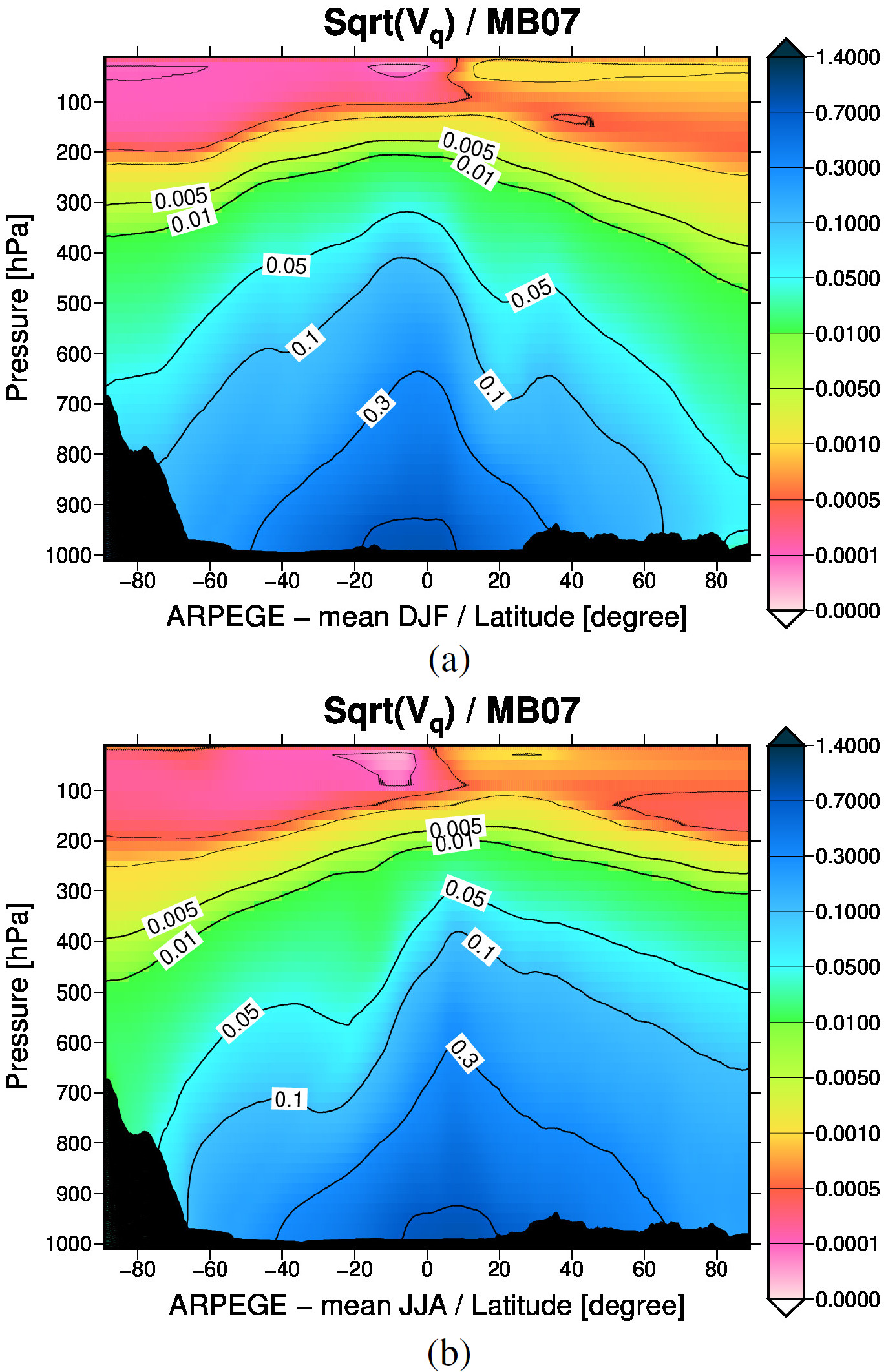}
\vspace*{-5mm}
\caption{\it \small 
The same as Figs.~\ref{fig_ARP_DJF_JJA_water_incr_anal_exe}~(b) and (d),
but for the DJF and JJA seasonal average of the MB07 water term $\sqrt{V_{q2}}$ given by (\ref{def_N99_def_Vq2}).
\label{fig_ARP_DJF_JJA_water_sqrt_MB07}}
\end{figure}

\begin{figure*}[hbt]
\centering
\includegraphics[width=0.99\linewidth]{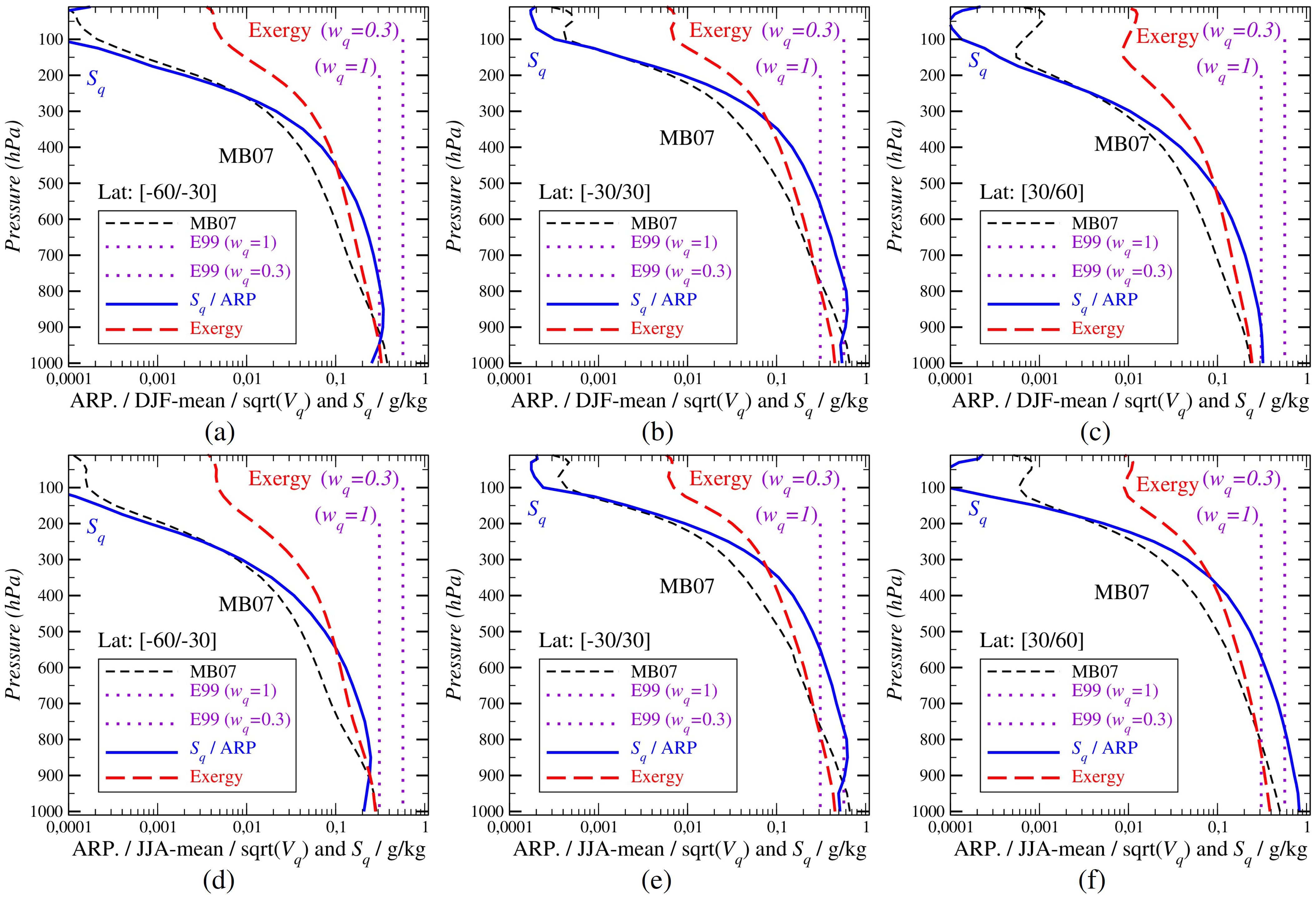}
\vspace*{-3mm}
\caption{\it \small 
Vertical profiles of horizontal mean of seasonal averages computed from ARPEGE outputs every $6$~hours and for three latitude domains: 
(a) and (d) southern extra-tropical mid-latitudes from $-60$ to $-30$ degrees;
(b) and (e) tropical latitudes from $-30$ to $+30$ degrees;
(c) and (f) northern extra-tropical mid-latitudes from $+30$ to $+60$ degrees.
The vertical profiles of the DJF means are plotted in (a, b, c); those for the JJA means in (d, e, f).
The E99 water terms $\sqrt{V_{q1}}$ (purple dotted lines) are given by (\ref{def_N99_def_Vq}) with $w_q=1.0$ and $w_q=0.3$.
The exergy water term $\sqrt{V_{q}}$ (red dashed lines) is given by (\ref{def_Nm_def_q}). 
The MB07 water term $\sqrt{V_{q2}}$ (black dashed lines) is given by (\ref{def_N99_def_Vq3}).
The RMS of analysis increments in water vapor is $S_q$ (blue solid lines).
\label{fig_ARPEGE_profiles_qv}}
\end{figure*}

The JJA and DJF seasonal means of the ``constant RH'' value $\sqrt{V_{q2}}$ derived in MB07 are shown in Fig.~\ref{fig_ARP_DJF_JJA_water_sqrt_MB07}.
The seasonal latitude oscillation is similar to that of $S_q$ and $\sqrt{V_{q}}$ in Fig.~\ref{fig_ARP_DJF_JJA_water_incr_anal_exe}.
The decrease with height of $\sqrt{V_{q2}}$ is larger than for the exergy norm, due to the property $w_{q2} \approx (w_{q})^2$ derived from (\ref{def_wz2})-(\ref{def_wz_MB07}) leading to values of $\sqrt{V_{q2}}$ smaller than $0.0005$~g~kg${}^{-1}$ in the stratosphere (purple color).
These values of $\sqrt{V_{q2}}$ are close to those for the RMS of analysis increments above the level $200$~hPa.

Vertical profiles are plotted in Fig.~\ref{fig_ARPEGE_profiles_qv} for the horizontal means of the RMS of analysis increment $S_q$ and for the $V$-terms $\sqrt{V_{q}}$ (exergy), $\sqrt{V_{q1}}$ (E99) and $\sqrt{V_{q2}}$ (MB07).

Almost the same features are observed for the two seasons and for the three latitude domains.
The large decrease with height by at least 3 orders of magnitude for the analysis increments $S_q$ cannot be represented by the E99 constant values $\sqrt{V_{q1}} \approx 0.31$ or $0.57$~g~kg${}^{-1}$ with $w_q=1.0$ or $w_q=0.3$, nor for any other constant value for $w_q$.

The differences between the vertical profiles of the RMS of analysis increments, those for the exergy terms and those for the MB07 term remain small from the surface up to about $200$~hPa (less than one order of magnitude).
The exergy term $\sqrt{V_{q}}$ is almost similar to the RMS of analysis increments for the layer $500$-$250$~hPa in the tropical and summer extra-tropical regions, with the blue and red lines intersecting each other.
For levels above $100$~hPa, the MB07 term is closer to the RMS of analysis increments than the exergy term, with a rapid decrease with height that the exergy term cannot reproduce.

For these reasons, the RMS of the analysis increments, the exergy norm and the MB07 norm are thus similar to each other, while the values for E99 are more different from the other three.
The aim was not to perfectly simulate the RMS of the analysis increments, but to approach them qualitatively, both for their vertical variation and for their order of magnitude.

The lack of a contribution from condensed water species to the moist-air exergy norm, together with the absence of any latent heat terms $L_v$ or $L_s$, may seem surprising. 
However, the condensed water contents $q_l$ and $q_i$ do exist in (\ref{defam}) for the moist-air exergy function $a_m$, which forms the starting point for deriving the moist exergy squared norm.

It is this theory that ultimately allows $q_l$ and $q_i$ to be neglected in the squared norm components $N_T$, $N_p$ and $N_v$, as small correction terms.
Moreover, the seasonal averages plotted in Fig.~\ref{fig_ARP_DJF_JJA_water_incr_anal_exe} for  ARPEGE confirm that there is no need to add independent norms related to the condensates $q_l$ or $q_i$, because the comparisons between the latitude-section of $S_q$ and $\sqrt{V_{q}}$ do not reveal missing structures related to the convective regions where $q_l$ and $q_i$ are large (tropical cumulus and extra-tropical frontal regions).

 \subsection{\underline{Seasonal means of ARPEGE:} \\
             \underline{the temperature norms}.}
 \label{subsection_result_ARPEGE_tempe}

The exergy norm seemed able to induce new results, especially for the moisture term $\sqrt{V_q}$ due to the term ${\overline{r_v}(p)}$ in (\ref{def_Nm_def_q}), a result confirmed in the previous section.
Similarly, since the ratio ${\overline{T}(p)}/{T_r}$ in (\ref{def_Nm_def_Tp}) varies with pressure, and therefore with height, one may wonder whether this variation predicted by the theory is realistic or not.

\begin{figure}[hbt]
\centering
\includegraphics[width=0.99\linewidth]{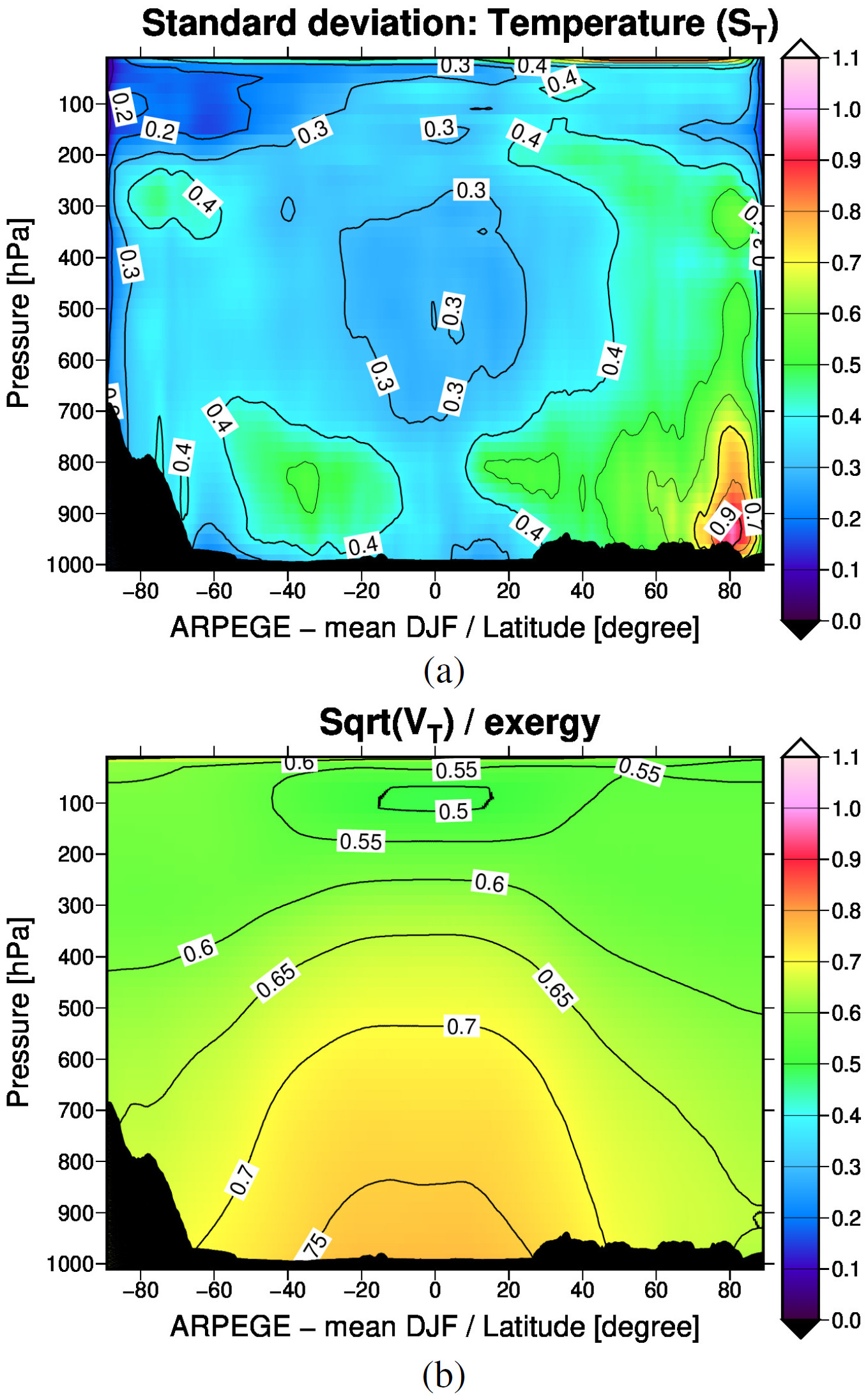}
\vspace*{-5mm}
\caption{\it \small 
The same as Figs.~\ref{fig_ARP_DJF_JJA_water_incr_anal_exe}a and \ref{fig_ARP_DJF_JJA_water_incr_anal_exe}b,
but for temperature (K).
\label{fig_ARP_DJF_JJA_T_incr_anal}}
\end{figure}

\begin{figure}[hbt]
\centering
\includegraphics[width=0.8\linewidth]{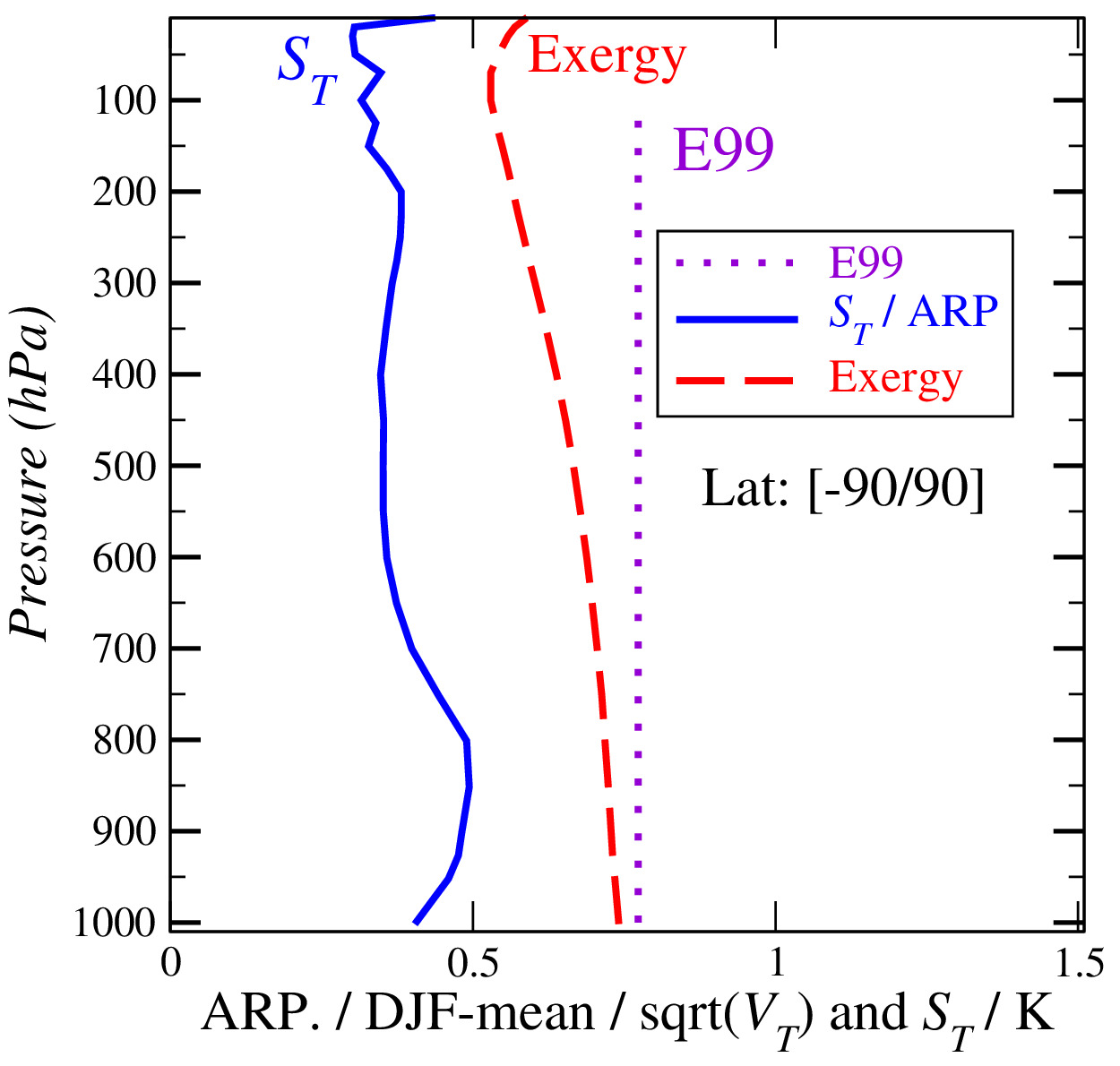}
\vspace*{-5mm}
\caption{\it \small 
The same ARPEGE seasonal mean (DJF) as in Figs.~\ref{fig_ARPEGE_profiles_qv}a
but for temperature (K) and for the RMS of analysis increments $S_T$ (solid blue), the E99 term $\sqrt{V_{T1}}\approx 0.77$ (dotted purple) and the exergy term $\sqrt{V_{T}}$ (red dashed).
\label{fig_ARPEGE_profiles_T}
}
\end{figure}

For this purpose, ARPEGE winter averages of the RMS of analysis increments for temperature $S_T$ and of the temperature exergy term $\sqrt{V_T}$ are shown in Fig.~\ref{fig_ARP_DJF_JJA_T_incr_anal}.
The corresponding vertical profiles of horizontal mean values are plotted in Fig.~\ref{fig_ARPEGE_profiles_T}.
The summer averages exhibit similar results (not shown).

Although the comparisons of norms for each latitude and pressure are less relevant for the temperature components than for the water components (especially within the tropics), the general appearance for $S_T$ and $\sqrt{V_T}$ is approximately similar, with a maximum near the surface (between $1000$ and $800$~hPa), and a minimum in the tropical troposphere for medium and high levels (between $600$ and $100$~hPa).

The variations with height of $\sqrt{V_T}$ are similar to those for $S_T$, while the constant value deduced from the E99 temperature component of the norm ($\sqrt{V_{T1}}\approx 0.77$ for $T_r=300$~K) is further from the $S_T$ profile.

Therefore, although variations with height of $S_T$ and $\sqrt{V_{T}}$ are smaller than those for $S_q$ and $\sqrt{V_{q}}$, the similarity between the vertical profiles of the seasonal averages of $S_T$ and $\sqrt{V_{T}}$ confirms the possible crude interpretation of the RMS of analysis increments with the temperature term computed from the squared exergy norm, and with a realistic impact for the ratio $({\overline{T}(p)}/{T_r})^2$ in (\ref{def_Nm_def_Tp}).

 \subsection{\underline{A specific day for CMC and GEOS} \\
             \underline{systems}.}
 \label{subsection_result_CMC_GEOS}

The results presented in the previous sections regarding ARPEGE seasonal averages are encouraging, but the need for daily applications of the exergy norm would require similar variations with height and  latitude for a given situation for both the analysis increments and the norms.
In addition, the encouraging results obtained with the 4D-Var incremental assimilation of the ARPEGE variable mesh model must be confirmed with different models and/or assimilation schemes.

\begin{figure*}[hbt]
\centering
\includegraphics[width=0.99\linewidth]{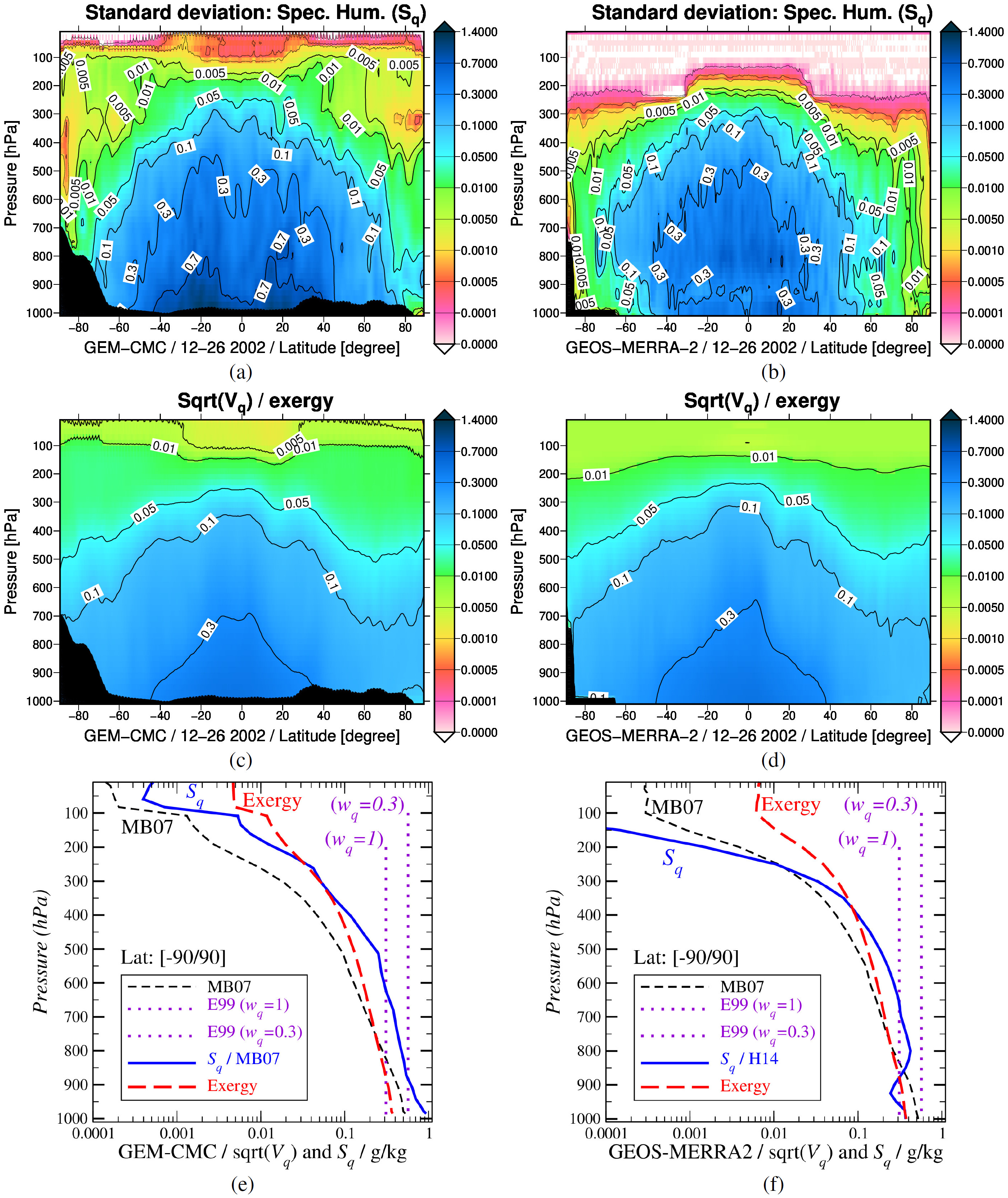}
\caption{\it \small 
Latitude-pressure sections and vertical profiles of horizontal averages for the water term for the 26th of December 2002: in the left panels (a, c, e) for the GEM-CMC; in the right panels (b, d, f) for GEOS-MERRA-2.
At the top (a, b): sections of the RMS of analysis increments $S_q$ (g~kg${}^{-1}$).
On the center (c, d): sections of exergy norms  $\sqrt{V_{q}}$ (g~kg${}^{-1}$).
At the bottom (e, f): vertical profiles of horizontal averages of E99 (dotted purple), MB07 (dashed black) and Exergy (dashed red) norms and the analysis increments $S_q$ (solid blue). 
\label{fig_CMC_GEOS_water_incr_anal_exe}}
\end{figure*}

To do this, the results obtained for the humidity variable are shown in Figs.~\ref{fig_CMC_GEOS_water_incr_anal_exe} for one single analysis (26 December 2002, 00 UTC).
Outputs from the GEM-CMC system are on the left in (a, c, e) and those from the GEOS-MERRA-2 system are on the right in (b, d, f).
The latitude-pressure sections for the RMS of analysis increments $S_q$ in (a, b) are similar to those in Figs.\ref{fig_ARP_DJF_JJA_water_incr_anal_exe}~(a, b).
The vertical profiles of the exergy term  $\sqrt{V_q}$, the MB07 term $\sqrt{V_{q2}}$ and the E99 terms $\sqrt{V_{q1}}$ computed with $T_r=300$~K and $w_q = 1.0$ or $w_q = 0.3$ are similar to those in Figs.\ref{fig_ARPEGE_profiles_qv}~a.

While the RMS of analysis increments are noisier for those GEM-CMC and GEOS-MERRA-2 daily outputs than for the ARPEGE seasonal averages, the same decay with height and relative maxima in the lower layers in the tropics is observed for this particular day.
The differences between the three ARPEGE, GEM-CMC and GEOS-MERRA-2 systems are more pronounced above $200$~hPa in the upper troposphere and in the stratosphere, where GEM-CMC exhibits larger analysis increments than ARPEGE, while those for GEOS-MERRA-2 are smaller than ARPEGE.

The latitude-pressure sections plotted for the water component of the exergy norm in Figs.~\ref{fig_CMC_GEOS_water_incr_anal_exe}~(c, d) for GEM-CMC and GEOS-MERRA-2 are similar to those for ARPEGE in Figs.\ref{fig_ARP_DJF_JJA_water_incr_anal_exe}~(a, b).

The water exergy $SqV$-term $\sqrt{V_{q}}$ is relatively smooth and not noisy because it depends on the averaged value of the water vapor $\overline{q_v}$ computed on a circle of latitude, which is less variable in space than the daily RMS of analysis increments $S_q$.

The results presented in this section for a specific day and for two different systems are therefore broadly comparable to those shown for the ARPEGE seasonal averages.
We can therefore be confident that the results derived in this paper from the exergy norm will be robust for other systems with similar patterns of analysis fields.

\newpage
\newpage

\clearpage

\begin{figure}[hbt]
\centering
\includegraphics[width=0.9\linewidth]{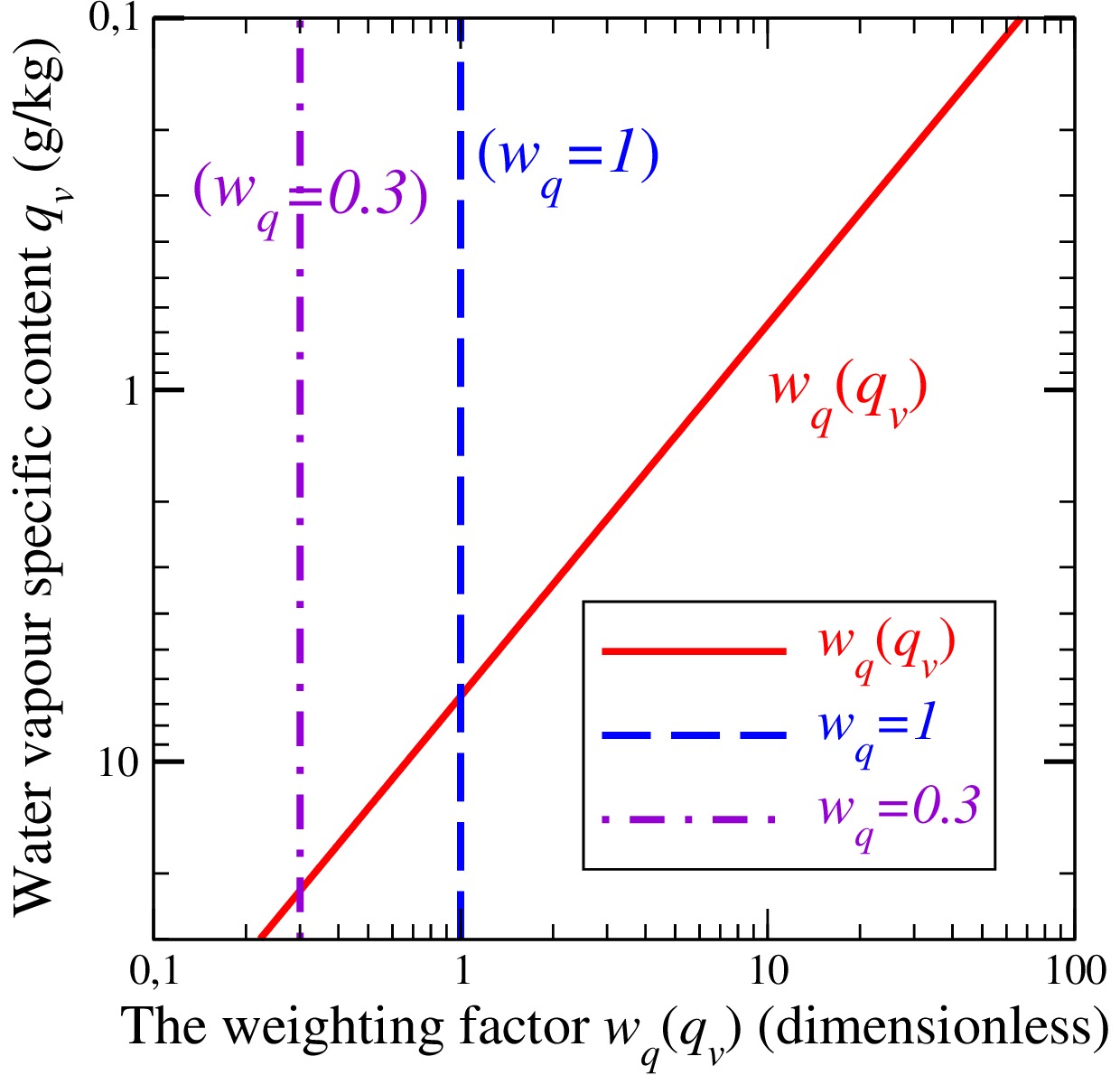}
\caption{\it \small 
The dimensionless exergy weighting factor $w_q(q_v)$ given by (\ref{def_wz}) plotted with $q_v$ in ordinates.
\label{fig_MB07_qv_wqB}}
\end{figure}

\begin{figure}[hbt]
\centering
\includegraphics[width=0.9\linewidth]{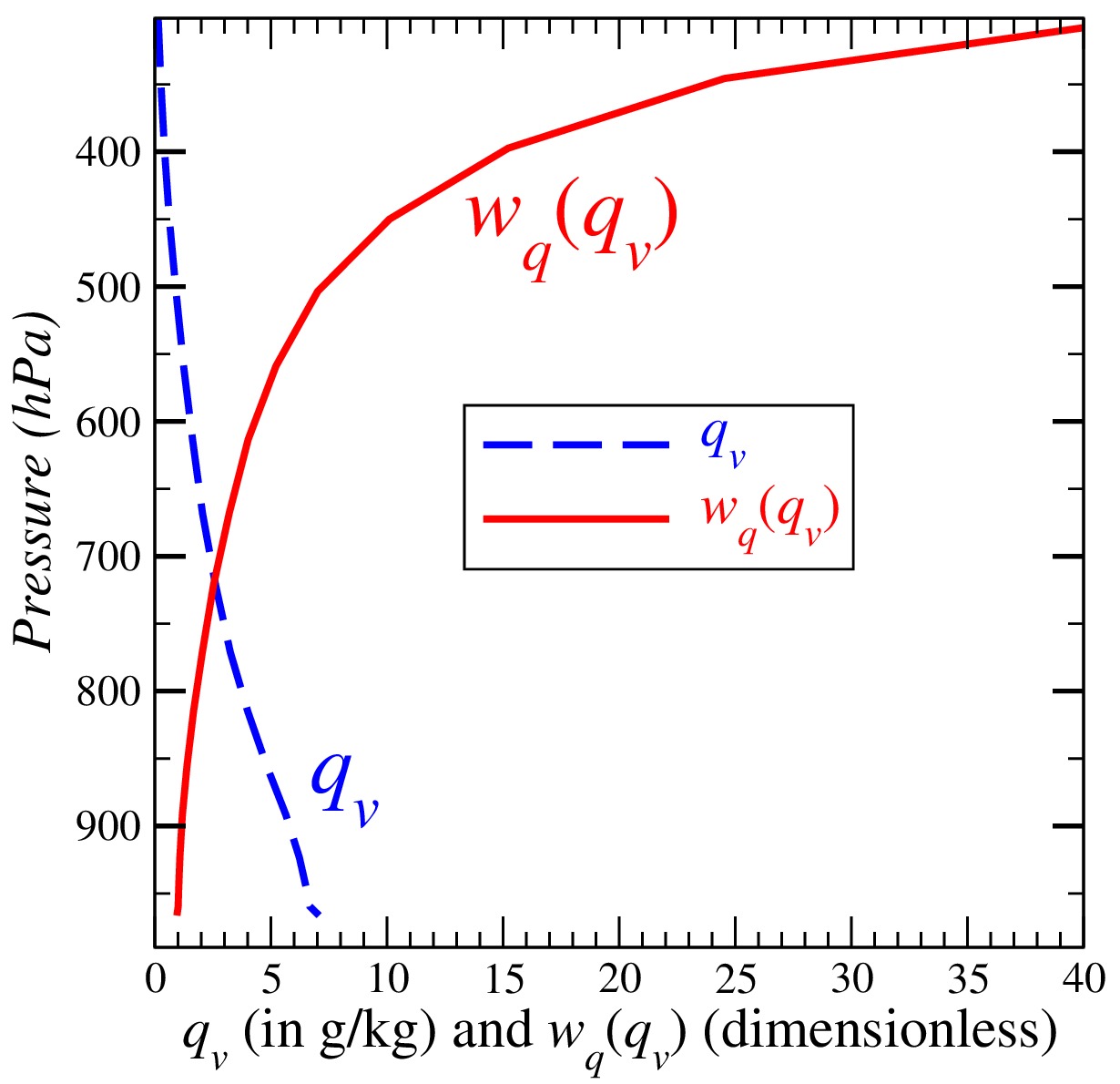}
\caption{\it \small 
The dimensionless exergy weighting factor $w_q(z)$ given by (\ref{def_wz}) for the vertical profile of average values $\overline{q_v}(p)$ of the GEM-CMC dataset used to plot the Fig.~\ref{fig_CMC_GEOS_water_incr_anal_exe}c.
\label{fig_MB07_qv_wqA}}
\end{figure}

\clearpage

 \subsection{\underline{The decrease with height of $w_q$}.}
 \label{subsection_result_decrease_wq}
 
The advantage of the exergy approach is that it provides an analytic formulation for the weighting factor $w_q$ given by (\ref{def_wz}).
As an example, values of $w_q(q_v)$ are plotted in Fig.~\ref{fig_MB07_qv_wqB} for $0.1 < q_v < 25$~g~kg${}^{-1}$.

The weighting factor $w_q(r_v)$ is smaller than unity for moist low levels where $q_v > 6.7$~g~kg${}^{-1}$ for $T_r=300$~K, and it is equal to $0.33$ for $q_v \approx 20$~g~kg${}^{-1}$. 
Conversely, it is much larger than unity for small values of $q_v$, reaching $w_q \approx 67$ for $q_v \approx 0.1$~g~kg${}^{-1}$ in the upper troposphere and $w_q \approx 6700$ for $q_v \approx 0.001$~g~kg${}^{-1}$ in the stratosphere.

It is also possible to plot the vertical profiles of $w_q$ in terms of the horizontal mean value $\overline{q_v}(p)$ computed from the GEM-CMC simulation, shown in Fig.~\ref{fig_MB07_qv_wqA}.
The large increase of $w_q$ with height, with a factor varying non-linearly from $1$ to $40$ for the pressure varying from $1000$~hPa to $300$~hPa, is similar to the one proposed empirically in previous studies;
for instance, a weight of $w_q(r_v) \approx 5$ was evaluated for the lower part of the atmosphere in \citet{Barkmeijer_al01} from the ECMWF averaged error variances for $q_v$, with $w_q(r_v)$ strongly increasing above $500$~hPa.
This description is consistent with the exergy weight displayed in Fig.~\ref{fig_MB07_qv_wqA}.

The same relation used to plot these diagrams ``$w_q$ in terms of $q_v$'' is used to plot the exergy norm in the pressure ($p$) and latitude ($\varphi$) sections shown in Figs.~\ref{fig_ARP_DJF_JJA_water_incr_anal_exe}~(b and d) and \ref{fig_CMC_GEOS_water_incr_anal_exe}~(c and d), where the zonal averages $\overline{q_v}(\varphi,p)$ varies with both pressure and latitude.

\begin{table*}
\caption{\it \small 
The increase in observation impacts (in percentage) corresponding to Fig.~\ref{fig_H17_FSOI_obs_impact} for the change of the Dry norms to the 
 moist E99 with $w_q=0.3$ (first line), and then to the moist Exergy (second line).
\vspace*{1mm}
\label{Table_FSOI}}
\centering
\begin{tabular}{|c|c|c|c|c|c|c|c|}
\hline 
              & AMSU-A & IASI  & {\bf MHS}    & AIRS  &  AMVs & RAOBs & Aicrafts \\ 
\hline 
 $100 \: (\: \mbox{E99 / Dry}-1 \: )$     
              & $14$ & $17$  & $300$     & $32$  & $21$  & $18$  & $10$ \\ 
 $100 \: (\: \mbox{Exergy / E99}-1 \: )$  
              & $47$ & $85$  & \boldmath{\mbox{$253$}} 
                                         & $105$ & $69$  & $74$  & $50$ \\
\hline 
\end{tabular}
\end{table*}

 \subsection{\underline{FSOI}.}
 \label{subsection_result_FSOI}

The Forecast Sensitivity to Observation Impact (FSOI) method can be used to assess and compare the capacity of various observing systems to reduce a given short-range forecast error produced by a NWP model, e.g. \citet{Baker_Daley00}, \citet{Langland_Baker_2004}, \citet{Cardinali_2009},  \citet{Gelaro_al_2010}. 
Typically, fields from a $24$~h forecast are compared against a verifying analysis, in terms of $u$, $v$, $T$, $p_s$ and $q_v$ using an inner product based on the E99 energy norm with different values of $w_q$ in the moist term. 
The adjoint of the forecast model is used to propagate a sensitivity backwards from verifying time ($24$~h) to obtain a sensitivity at analysis time ($0$~h). 
The adjoint model can include both dry physical processes (turbulent diffusion, radiation, gravity wave drag) and moist processes (large scale condensation, moist convection).

Impacts shown in the present paper are examined in averages per observation system and for the global domain with the E99 norms (\ref{def_N99_def_VuTp})--(\ref{def_N99_def_Vq}) where $T_r=270$~K, $p_r=1000$~hPa and $w_q = 0.3$.
The value of $0.3$ is chosen empirically in H14 to produce approximately equal weighting between the temperature and specific humidity components of the norm.

The metrics monitored at GMAO are: impact per analysis, impact per observation, fraction of beneficial observations, and observation count per analysis.
The observation impacts are computed as reductions in the final $24$~h forecast errors due to any given extra set of observations included in the initial analysis.
The adjoint model can be used to propagate the final energy norm gradient backward $24$~h in order to obtain sensitivities of these forecast errors at the initial time  \citep{Tremolet_2008}.
These sensitivities are then passed through the adjoint of the data assimilation system to convert them into observation space and to provide the impacts.

\begin{figure}[hbt]
\centering
\includegraphics[width=0.99\linewidth]{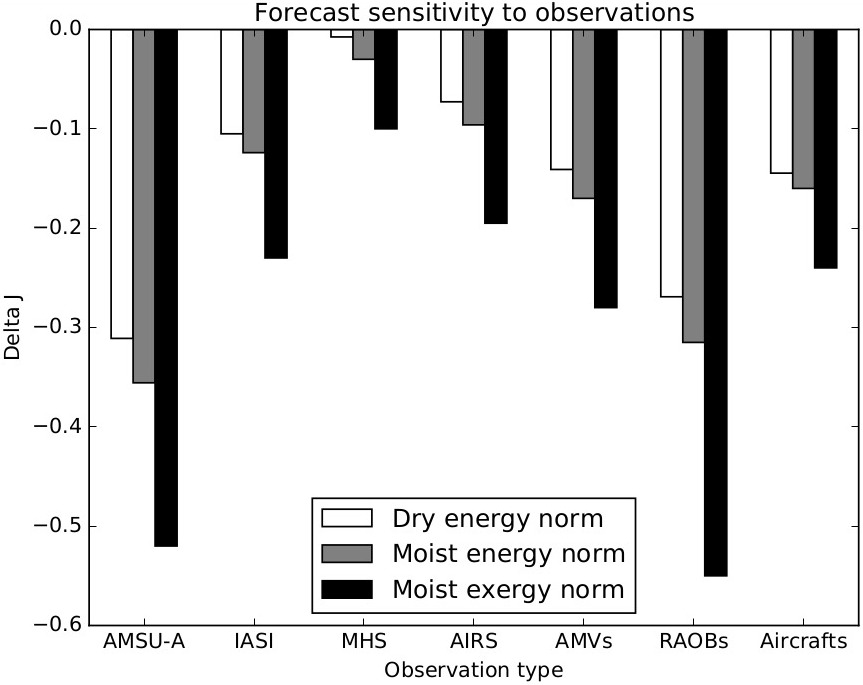}
\caption{\it \small 
The 24-h forecast observation impacts per analysis for each observation system.
Comparisons of:
i) the Dry norm (white);
ii) the moist norm E99  with $w_q = 0.3$ (grey), namely the same as Fig.~9 in H14;
and iii) the moist exergy norm (dark).
\label{fig_H17_FSOI_obs_impact}}
\end{figure}

Fig.~\ref{fig_H17_FSOI_obs_impact} compares the $24$~h forecast error reductions produced by various observing systems included in the MERRA-2 data assimilation system with three different inner products for the estimation of the global forecast error: the E99 ``dry energy squared norm'' with $w_q=0.0$, the E99 ``moist energy squared norm'' with $w_q=0.3$, and the ``exergy squared norm'' $N_T+N_p+N_v$ introduced in Eqs.~(\ref{def_Nm_T})-(\ref{def_Nm_v}) of 
Section~\ref{subsection_theory_new_norm}. 

The impacts of the dry energy E99 squared norms are those computed and studied in Fig.9 of H14 for the month 17 March-17 April 2012.
The impacts for the two moist squared norms (E99 with $w_q=0.3$ and exergy formulations) are computed for another month (1-30 September 2015).
For convenience, the impacts of the three dry and moist squared norms are compared on the same plot despite having been computed over those two distinct periods.
In all experiments, the adjoint model includes a comprehensive set of physical processes with moist processes as described in H14. 

As expected from the definition of the moist energy norm, impacts are larger when they include the moist term, as already shown in H14. 
It is interesting to note that the increase in observation impacts not only holds for observations sensitive to atmospheric water vapor, such as radiosoundings, but also for observation systems where only a small subset of the observations directly measure moisture, such as IASI (Infrared Atmosphere Sounding Interferometer) radiances, AMSU-A (Advanced Microwave Sounding Unit) radiances which are sensitive to atmospheric temperature, and AMVs (Atmospheric Motion Vectors) which are directly sensitive to horizontal wind components. 
These results show that a reduction of forecast error in the moisture field is possible through observations of temperature and wind.
This could occur through dynamical balance, for example.

The ranking, in terms of contributions of the various observing systems to the forecast error reduction, is unchanged when moving from E99/$w_q=0.0$ to E99/$w_q=0.3$.
Similarly, when examining the impact with the exergy  norm instead, it is clear that the overall observation impact is larger, but that the ranking of the  observation systems relative to each other is almost the same.
Larger values come from the difference in the weighting factor $w_q$ applied to the moisture at upper levels, which does not depend on height for the E99 norm and increases with height for the exergy norm according to Fig.~\ref{fig_MB07_qv_wqA}.

The most striking feature, when using the exergy norm, is the very large increase by a factor of three (or $> + 200$~\%, see Table.\ref{Table_FSOI}) of the only observing system highly sensitive to atmospheric water vapor: the Microwave Humidity Sounder (MHS). 
According to Fig.~\ref{fig_H17_FSOI_obs_impact}, RAOB ranks first for the exergy norm, which may have important implications given that the operational radiosonde observing network is expensive to operate.
These results suggest that radiosonde humidity sensors play an important role in the $24$~h forecast accuracy, even more than MHS.

\section{\underline{Conclusions}.} 
\label{section_conclusion}

The main objective of this paper is to provide a general and more satisfactory method for combining thermodynamic variables of the atmosphere into a norm. 
There are several formulations for these norms currently in use for a wide variety of important applications, yet until now all have been derived using heuristic methods and approximations.

It is argued in this paper that such approximations can be avoided by instead considering the principles of fundamental physics more carefully.
Specifically, the approach is to start with some general exergy functions, which are constructed by combining the first (enthalpy) and second (entropy) law of thermodynamics, leading to the available enthalpy function $a_m$ derived in M93. 
This kind of exergy function is also based on the concept of relative entropy or Kullback distance, two equivalent concepts which are already used in many papers dealing with assimilation techniques.

The choice of the exergy (available enthalpy) squared norms provides not only the quadratic terms $(T')^2$, $(p'_s)^2$ and $(q'_v)^2$, but also values for the weighting factors which multiply these quadratic terms.
It is shown in the present paper that the weighting factors for $T$ and $q_v$ vary with height in the same way as the RMS of analysis increments.
This ensures an even weighting of all variables and all levels when computing the global norm.
Such results are valid for both seasonal average periods and for a particular day.

The fact that the weights for the exergy norm for $T$ and $q_v$ are close to the RMS of analysis increments is not straightforward.
Indeed, if the observation system is radically changed, the increments could be very different, while the exergy-norm weights would not be modified.
To better understand the complex links that can exist between fields as different as thermodynamics, information theory and data assimilation, it is possible to refer to papers cited in section~3 of \citet{Marquet_Thibaut_2018}.

Inspired by previous studies by \citet{Kleeman_2002} and \citet{Majda_al_2002}, the paper of \citet{Xu_2006} examined the use of the relative entropy or Kullback-Leibler distance $K(x||y)$ given by (\ref{eq_K})
``{\it to measure the information content of the pdf produced by an optimal analysis of observations (or compressed super-observations) with respect to a prior background pdf used by the analysis (...) where the background pdf can be always considered as an approximation of the analysis pdf.}''
\citeauthor{Xu_2006} showed that the integral form of the relative entropy $K(x||y)$ ``{\it is a quadratic form of the analysis increment vector weighted by $\boldmath{B}^{-1}$}'', and ``{\it yields an explicit formulation in which the signal part is given by the inner-product of the analysis increment vector weighted by the inverse of the background covariance matrix}'' ($\boldmath{B}^{-1}$).

Since \citet{Xu_2006} demonstrated a close relationship between $K(x||y)$ and the weighting factors $V_u$, $V_v$, $V_v$, $V_p$, $V_T$ and $V_q$, the next step is to use the close relationship shown by \citet{Procaccia_Levine_1976}, \citet{Eriksson_Lindgren_1987}, \citet{Eriksson_al_1987}, \cite{Karlsson_90} and \citet{Honerkamp_1998} between $K(x||y)$ and exergy functions, to foresee a direct link between the moist-air exergy defined in thermodynamics and the weighting factors used in data assimilation.

The new exergy (available enthalpy) squared norm may solve the main disadvantage of using the constant E99 moist $V$-term stated in \citet{Riviere_al_2009}, namely that the weight for water is no longer proportional to the weight for temperature with the exergy formulation, leading to new results with the use of the $V_q$ term.

A first usage of the exergy norm in the context of FSOI experiments has shown that it increases observation impact in a way similar to what has previously been described when going from a dry energy norm to a moist energy norm (e.g. H14). 
However, the enhancement of the impact is larger, since the exergy norm accounts more evenly for moisture forecast errors between the various atmospheric layers, whereas the moist energy norm penalizes the upper tropospheric levels. 
The results are very similar among the various observing systems, however with a noticeable difference for the MHS and RAOBs, for which the contributions are particularly enhanced with the exergy norm. 
This is in agreement with the known impact of microwave humidity sounders from direct observing system experiments \citep{Karbou_al_2010,Chambon_al_2015}. 
In consequence, it is expected that the various observing systems would be more fairly ranked through more balanced contributions between wind, temperature and water vapor forecast errors through the use of the exergy norm in FSOI experiments.

Another usage of the exergy norm has been shown by \citet{Borderies_2019} to demonstrate the impact of airborne cloud radar reflectivity data assimilation.

The important point is that the analytical formulation of the exergy norm is not complicated.
It is comparable in complexity to existing formulations (E99, MB07) and can be easily coded and used in operational systems, for moist singular vector and FSOI calculations as well as forecast verifications.
The only new aspect is the need to take into account horizontal averages, or averages on each latitude circle, for the mean temperature and vapor content variables $\overline{T}$ and $\overline{r_v}$ that appear in (\ref{def_Nm_T})-(\ref{def_Nm_def_q}) to define $N_T$, $N_v$, $V_T$ and $V_q$.


\vspace{2mm}
{\large \bf \underline{Acknowledgements}.}
\vspace{2mm}

The definitions of the squared norm components $N_T$, $N_p$ and $N_v$ were obtained during the Pan-GCSS meeting in Athens in May 2005.
The results presented in this paper are thanks to Philippe Courtier's initial encouragements, with numerous preliminary tests carried out between 2005 and 2018.
The authors wish to thank the editor and the three reviewers for their comments, which helped to improve the manuscript.

\clearpage


\vspace{4mm}
\noindent
{\bf\underline{Appendix~A. List of symbols and acronyms}.}
                             \label{appendixA}
\renewcommand{\theequation}{A-\arabic{equation}}
  \renewcommand{\thefigure}{A-\arabic{figure}}
   \renewcommand{\thetable}{A-\arabic{table}}
      \setcounter{equation}{0}
        \setcounter{figure}{0}
         \setcounter{table}{0}

\begin{tabbing}
 ----------\=  --------------------------------- ---\= \kill
 $B_p$        \> a dummy notation for a pressure norm \\
 $APE$        \> the global available potential energy (Lorenz)\\
 $\alpha$ \> the specific mass of moist air (the density $1/\rho$) \\ 
 $a_e$        \> the moist specific available energy \\
 $a_h$, $a_m$ \> the dry and moist specific available enthalpies \\
 $a_T$, $a_p$ \> temperature and pressure components of \\
              \> $a_h$ and $a_m$ \\
 $a_v$        \> the water component of $a_m$ \\
 $c_{pd}$ \> specific heat of dry air ($1004.7$~J~K${}^{-1}$~kg${}^{-1}$) \\
 $c_{pv}$ \> spec. heat of water vapor ($1846.1$~J~K${}^{-1}$~kg${}^{-1}$) \\
 $c_{l}$  \> spec. heat of liquid water ($4218$~J~K${}^{-1}$~kg${}^{-1}$) \\
 $c_{i}$  \> spec. heat of ice      ($2106$~J~K${}^{-1}$~kg${}^{-1}$) \\
 $c_p$ \> the moist-air spec. heat at constant pressure, \\
       \> $ = \: q_d \: c_{pd} + q_v \: c_{pv} + q_l \: c_l + q_i \: c_i  $ \\
 $\delta$ \> $=R_v/R_d-1 \approx 0.608$ \\
 $e$        \> the water-vapor partial pressure \\
 $e_i$      \> the specific internal energy \\
 $e_r$      \> the water-vapor reference partial pressure,\\
            \> with $\: e_r = e_{sw}(T_r) \approx 6.11$~hPa \\
 $F$, $H$ \> dimensionless functions of $X$ or $Y$ \\
 $g$  \> magnitude of Earth's gravity ($9.8065$~m${}^{2}$~s${}^{-2}$)\\
 $\overline{\Gamma}$ \> the Lorenz stability parameter \\
 $\overline{\Gamma}_q$ \> a weight in water component of MB07 norm \\
 $h$, $H$ \> specific and global enthalpies\\
 $H_r$   \> a dummy scale height (C87) \\
 $k_B$   \> the Boltzmann constant \\
 $K$ \> Kullback function, contrast, relative entropy \\
 $L_f$    \> $=h_l-h_i$: latent heat of melting \\
 $L_v$    \> $=h_v-h_l$: latent heat of vaporization \\
 $L_s$    \> $=h_v-h_i$: latent heat of sublimation \\
 $L_f(T_r)$  \> $= 0.334$~$10^{6}$~J~kg${}^{-1}$\\
 $L_v(T_r)$  \> $= 2.501$~$10^{6}$~J~kg${}^{-1}$\\
 $L_s(T_r)$  \> $= 2.835$~$10^{6}$~J~kg${}^{-1}$\\
 $m$      \> a mass of moist air \\
 $dm$     \> the element of mass ($= \rho \: d\tau$) \\
 $N$      \> the global available enthalpy squared norms \\
 $\omega_{ij}$ \> the fractional coverage of the model grid box \\
 $x_j, y_j$ \> the micro states which define the function $K$  \\
 $p$      \> the pressure ($p= p_d + e$)  \\
 $p_s$    \> the surface pressure \\
 $q$   \> the specific content (ex. $q_v={\rho}_v / {\rho}$) \\
 $Q_r$   \> a dummy specific water content (C87) \\
 $r$   \> the mixing ratio (ex. $r_v={\rho}_v / {\rho}_d$)\\
 $r_0$ \> $=R_d/R_v \approx 0.622 = 1/1.608$ \\
 ${\rho}$   \> specific mass of  moist air  
              ($={\rho}_d+{\rho}_v+{\rho}_l+{\rho}_i$)  \\
 $R_d$   \> dry-air gas constant ($287.06$~J~K${}^{-1}$~kg${}^{-1}$) \\
 $R_v$   \> water-vapor gas constant ($461.52$~J~K${}^{-1}$~kg${}^{-1}$) \\
 $R$     \> gas constant for moist air ($ = q_d \: R_d + q_v \: R_v$)\\
 $s$ \> the specific entropy \\
 $\sigma$ \> the vertical coordinate of the model grid box \\
 $\Sigma$, $d\Sigma$ \> global and element of horizontal surface of Earth \\
 $T$       \> the absolute temperature \\
 $T_{r}$   \> the reference zero Celsius temperature ($273.15$~K) \\
$\boldsymbol{U}$ \> the horizontal wind and its components $(u, v)$ \\
$U$ \> the horizontal wind speed $\sqrt{u^2 + v^2}$ \\
${\mu}$ \>  the specific Gibbs' function ($h-T\:s$) \\
$\phi$  \>  the gravitational potential energy ($g\: z + Cste$ ) \\
$V$     \> the variances of analysis errors \\
$V_0$   \> a special variance of $2$~J~kg${}^{-1}$ \\
$w_q$     \> a relative weight in water components of norms \\
$Z$     \> a dimensionless water vapor variable \\
GCM   \> General Circulation Model \\
NWP   \> Numerical Weather Prediction \\
 \=  --------------------------------- --\= \kill
  \> \\
  \>  \underline{Lower indices} (for $h$, $s$, $p$, $\mu$, 
  $\rho$, $q$, $r$, $V$, $X$, $Y$, $Z$): \\
 ----------\=  --------------------------------- --\= \kill
 $r$ \> reference value (entropy, available enthalpy) \\
 $d$, $v$  \> dry-air and water vapor gases phases \\
 $l$, $i$  \> liquid water and ice condensed phases \\
 $sw$, $si$  \> saturating value (over liquid or ice) \\
 $t$ \> total water value (vapor plus liquid plus ice) \\
 $T$, $p$, $v$ \> temperature, pressure and water components \\
 $T_1$, $p1$ \> notations for pressure components ($V$) \\
 $q$, $q2$ \> notations for water components ($V$) \\
 $1$, $2$ \> notations in separating laws \\
 $i, j, k$ \> indices for longitude, latitude and altitude \\
 \=  --------------------------------- --\= \kill
  \> \\
  \>  \underline{Upper indices/operator}: \\
 ----------\=  --------------------------------- --\= \kill
 $(\ldots)'$ \>  departure terms from average values \\ 
 $\overline{(\ldots)}$ \> average values
\end{tabbing}

\vspace{4mm}
\noindent
{\bf \underline{App.~B. The specific moist-air available enthalpy}.}
                           \label{appendixB}
\renewcommand{\theequation}{B-\arabic{equation}}
  \renewcommand{\thefigure}{B-\arabic{figure}}
   \renewcommand{\thetable}{B-\arabic{table}}
      \setcounter{equation}{0}
        \setcounter{figure}{0}
         \setcounter{table}{0}
\label{---Appendix-B}
\vspace{2mm}

The specific moist available enthalpy $a_m$ is an exergy function defined in M93 (see Eq.~(17), page 574) as a sum of four partial moist available enthalpies for dry air $(a_m)_d$, water vapor $(a_m)_v$, liquid water $(a_m)_l$ and ice $(a_m)_i$, leading to
\vspace{-0.15cm}
\begin{align}
\!\!\!\!
  a_m  & = q_d \: (a_m)_d + q_v \: (a_m)_v 
          + q_l \: (a_m)_l + q_i \: (a_m)_i \: , \label{defam}\\
\!\!\!\!
     & (a_m)_d  = [ \: h_d - (h_d)_r \: ] \: - \: T_r \: [ \: s_d - (s_d)_r \: ] \: , \label{defamd}\\
\!\!\!\!
     & (a_m)_v  = [ \: h_v - (h_v)_r \: ] \: - \: T_r \: [ \: s_v - (s_v)_r \: ] \: , \label{defamv}\\
\!\!\!\!
     & (a_m)_l  = [ \: h_l - (h_l)_r \: ] \: - \: T_r \: [ \: s_l - (s_l)_r \: ] \: , \label{defaml}\\
\!\!\!\!
     & (a_m)_i = [ \: h_i - (h_i)_r \: ] \: - \: T_r \: [ \: s_i - (s_i)_r \: ] \: ,
\label{defami}
\end{align}
where $T_r$ is a constant reference pressure.

Differences in enthalpy and in entropy can be computed for dry air, water vapor and condensed species by assuming that the specific heat at constant pressure  ($c_{pd}$, $c_{pv}$, $c_l$, $c_i$) and gas constants ($R_d$, $R_v$) are all constant for the atmospheric range of temperature (from $180$ to $320$~K), leading to
\vspace{-0.15cm}
\begin{align}
\!\! \!\! \!\!\! \!\!\! \!\!\! \!\!\!   h_d - (h_d)_r  & =
      c_{pd} \: ( T - T_r )
      ,
     \hspace{1mm}
  h_v - (h_v)_r  =
      c_{pv} \: ( T - T_r )
      , \!\!\!  \label{defhdv} \\
\hspace{-4mm}
\!\! \!\! \!\!\! \!\!\! \!\!\! \!\!\!   h_l - (h_l)_r  & =
      c_l \: ( T - T_r )
      ,
     \hspace{1mm}
  h_i - (h_i)_r   =
      c_i \: ( T - T_r )
     , \!\!\!  \label{defhli}
\end{align}
and
\vspace{-0.15cm}
\begin{align}
\hspace*{-7mm}
  s_d - (s_d)_r  & =
      c_{pd} \ln( T / T_r ) - R_d \ln[ \: p_d / (p_d)_r \, ] 
     \: , \label{defsd} \\ 
\hspace*{-7mm}
  s_v - (s_v)_r  & =
      c_{pv} \ln( T / T_r ) - R_v \ln[ \: e / e_r \, ] 
     \: , \label{defsv} \\ 
\hspace*{-7mm}
  s_l - (s_l)_r  & =
      c_l \ln( T / T_r )
     \: ,
     \hspace*{1mm}
  s_i - (s_i)_r   =
      c_i \ln( T / T_r ) 
     \: . \!\!\! \label{defsli}
\end{align}
The reference partial pressure $e_r$ is equal to the ice-vapor value $e_{si}(T_r)$ for $T_r<0\,{}^{\circ}$~C or to the liquid-vapor value $e_{sw}(T_r)$ for $T_r>0\,{}^{\circ}$~C.
The moist available enthalpy (\ref{defam}) is computed by including (\ref{defhdv})-(\ref{defsli}) in (\ref{defamd})-(\ref{defami}), yielding
\vspace{-0.15cm}
\begin{align}
  \! \! \!  
  & a_m = \; c_p \left[ \:T -T_r - T_r \ln\!\left(\frac{T}{T_r} \right) \right]
  \nonumber \\
  \! \! \! 
   & + \: T_r \left [ \:
         q_d \: R_d \: \ln\!\left( \frac{p_d}{(p_d)_r} \right)
         + \:
        q_v \: R_v \: \ln\!\left( \frac{e}{e_r} \right) 
         \right]  .
\label{defam1}
\end{align}
Here $q_l$ and $q_i$ are not neglected, but appear in the moist values of $c_p$ and $q_d=1-q_v-q_l-q_i$, anywhere else.

\vspace{4mm}
\noindent
{\bf \underline{App.~C. The temperature component of $a_m$}.}
                            \label{appendixC}
\renewcommand{\theequation}{C-\arabic{equation}}
  \renewcommand{\thefigure}{C-\arabic{figure}}
   \renewcommand{\thetable}{C-\arabic{table}}
      \setcounter{equation}{0}
        \setcounter{figure}{0}
         \setcounter{table}{0}
\label{---Appendix-C}
\vspace{2mm}

The first term on the R.H.S. of (\ref{defam1}) is the Motivity defined by Lord Kelvin \citep{Thomson_1853}.
It corresponds to the moist temperature component $a_T$ of the available enthalpy defined in \citet[hereafter M91]{Marquet91} and M93 in terms of the function $F(X)$ according to
\vspace{-0.15cm}
\begin{align}
  a_T \, (T)   & = \:  c_p \;\; T_r \;\: F( X_T )  
 \: , \; \;
  X_T  \:  = \: T/\,T_r - 1  \; > \: -1 \: ,
  \label{def_at_Xt} \\
  F(X)  & = \:  X - \ln (1+X) \: .
  \label{def_FX}
\end{align}
The difference from the dry case studied in M91 is that $c_p$ is equal to $q_d \: c_{pd} + q_v \: c_{pv} + q_l \: c_l + q_i \: c_i$ and is not a constant, since it depends on varying specific contents of dry air and water species.
\begin{figure}[hbt]
\centering
\includegraphics[width=0.8\linewidth]{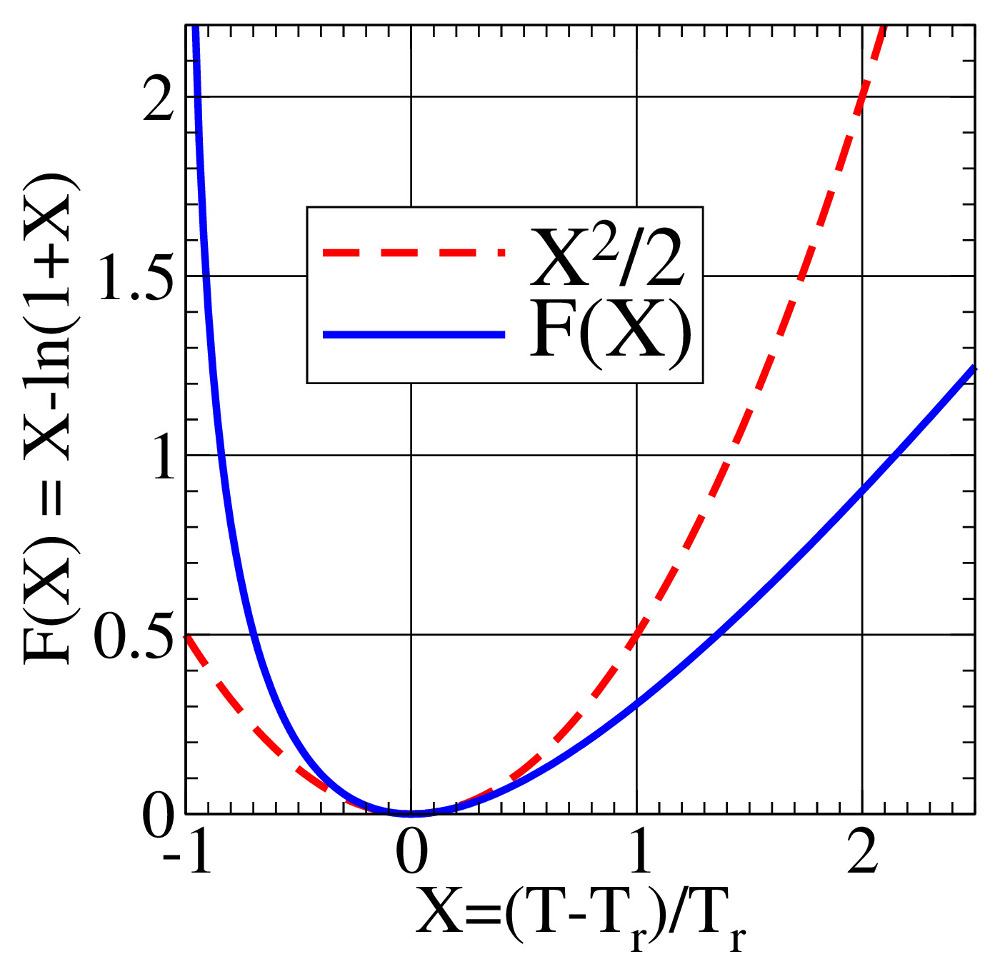}
\caption{\it \small 
The two functions 
$F(X)=X-\ln(1+X)$ 
and $X^2/2$ plotted for $-1<X<+2.5$.
\label{figFX}}
\end{figure}

$F(X)$ is positive and asymmetric with respect to $X=0$, see Fig.~\ref{figFX}
It is a quadratic-like function because $F(X) \approx X^2/2$ for $|X| < 0.3$.
This terminology ``quadratic-like'' corresponds to functions with Taylor series of the form: $X^2/2 + a \: X^3 + b \: X^4 + ...$, where the quadratic term $X^2/2$ is the first order approximation and where the other higher-order terms can be discarded.
This approximation is typically valid  for $210$~K$\;  < T < 390$~K if $T_r = 300$~K.
$F(X)=0$ only for $X=0$, namely for $T=T_r$.


\vspace{4mm}
\noindent
{\bf \underline{Appendix~D. The pressure components of $a_m$}.}
                             \label{appendixD}
\renewcommand{\theequation}{D-\arabic{equation}}
  \renewcommand{\thefigure}{D-\arabic{figure}}
   \renewcommand{\thetable}{D-\arabic{table}}
      \setcounter{equation}{0}
        \setcounter{figure}{0}
         \setcounter{table}{0}
\vspace{2mm}

Terms in the second line of (\ref{defam1}) can be rearranged in order to compute the separate quadratic contributions due to total pressure $p=p_d+e$ on the one hand, and to water species contents ($q_v$, $q_l$ or $q_i$) on the other hand.

The three state functions for moist air, dry air and water vapor can be written as $p=\rho\:R\:T$, $p_d  = q_d \: \rho \:R_d \:T$ and $e = q_v \: \rho \:R_v \:T$, respectively, leading to
\vspace{-0.15cm}
\begin{align}
  T_r \: q_d \: R_d   \; & = \; p_d \: T_r / (\rho\:T) \; 
                      = \; R \:T_r \:p_d / p \: , \label{defaux1}\\
  T_r \: q_v \: R_v   \; & = \; e \: T_r / (\rho\:T) \;  \hspace{0.2cm} 
                      = \; R \:T_r \:e / p \: ,   \label{defaux2}
\end{align}
where the moist gas constant $R=q_d \: R_d + q_v \: R_v$ is not a constant since it varies with $q_d$ and $q_v$.

The terms $q_d \: R_d$ and $q_v \: R_v$ given by (\ref{defaux1}) and (\ref{defaux2}) can be inserted into (\ref{defam1}), yielding
\vspace{-0.15cm}
\begin{align}
\!\!\!\!\!\!
  a_m  & = \:  a_T 
     + R \; T_r 
      \left[ \:
         \frac{p_d}{p} \:
         \ln \!\left( \frac{p_d}{(p_d)_r} \right) 
      +
         \frac{e}{e_r} \:
         \ln \!\left( \frac{e}{e_r} \right) 
      \right] \! .
\!\!
\label{defam2}
\end{align}
The next step is to isolate the pressure component $a_p$ defined by (\ref{defam3_p3}), leading to the separation of $a_m$ into
\vspace{-0.15cm}
\begin{align}
 \! \!
   a_m &  = \: a_T \: + \: a_p \: + \: a_v
  \: , \; \;  \; 
   a_p \: = R \; T_r 
             \: \ln\!\left( \frac{p}{p_r} \right) ,
\!\!
   \label{defam3_p3} \\
 \! \!
    a_v &  = 
     R \: T_r
           \left[
              \: \frac{p_d}{p} \:
              \ln\!\left( \!
                 \frac{p_d}{p} \: \frac{p_r}{(p_d)_r} 
              \right)
               +
              \frac{e}{p} \:
             \ln\!\left(
                 \frac{e}{p} \: \frac{p_r}{e_r}
              \right)
            \right] \!,
\!\!
 \label{defav3} 
\end{align}
where the remaining terms grouped in (\ref{defav3}) form the water components $a_v$.

It is not possible to define directly a squared norm starting from the term $\ln(p/p_r)$, since it is negative for $p<p_r$.
This apparent drawback was already mentioned in M91 and M93.
However, it is possible to integrate by parts $a_p$ in (\ref{defam3_p3}) with respect to $p$, leading to
\vspace{-0.15cm}
\begin{align}
\!\!\!\!
  a_p  & = R \: T_r \: p_r \: 
       \frac{\partial}{\partial p} \!
        \left[ \: 
             \frac{p}{p_r} \:
                 \ln \!\left(\frac{p}{p_r} \!\right)
           - \left( \frac{p}{p_r} -  C \right)
        \: \right]
        . \label{def0ap1}
\end{align}
A new quadratic-like function $H(X)$ can be introduced by choosing the constant of integration $C=1$, yielding
\vspace{-0.15cm}
\begin{align}
 \!\!\!\!
  a_p   & =   R \;  T_r \; 
      \frac{\partial}{\partial p }
          \left[\;  p_r
              \:H(X_p)\: 
          \right] 
  \: , \;\;
  X_p  \:  = \: \frac{p}{p_r} - 1 \: ,
 \!\!
 \label{def_ap_Xp}
  \\
 \!\!\!\!
 H(X) \; & = \; (1+X)\:\ln(1+X) \: - \: X  \: , 
\label{def_HX}
\end{align}
where $X_p$ is the dimensionless pressure control variable.

\begin{figure}[hbt]
\centering
\includegraphics[width=0.80\linewidth]{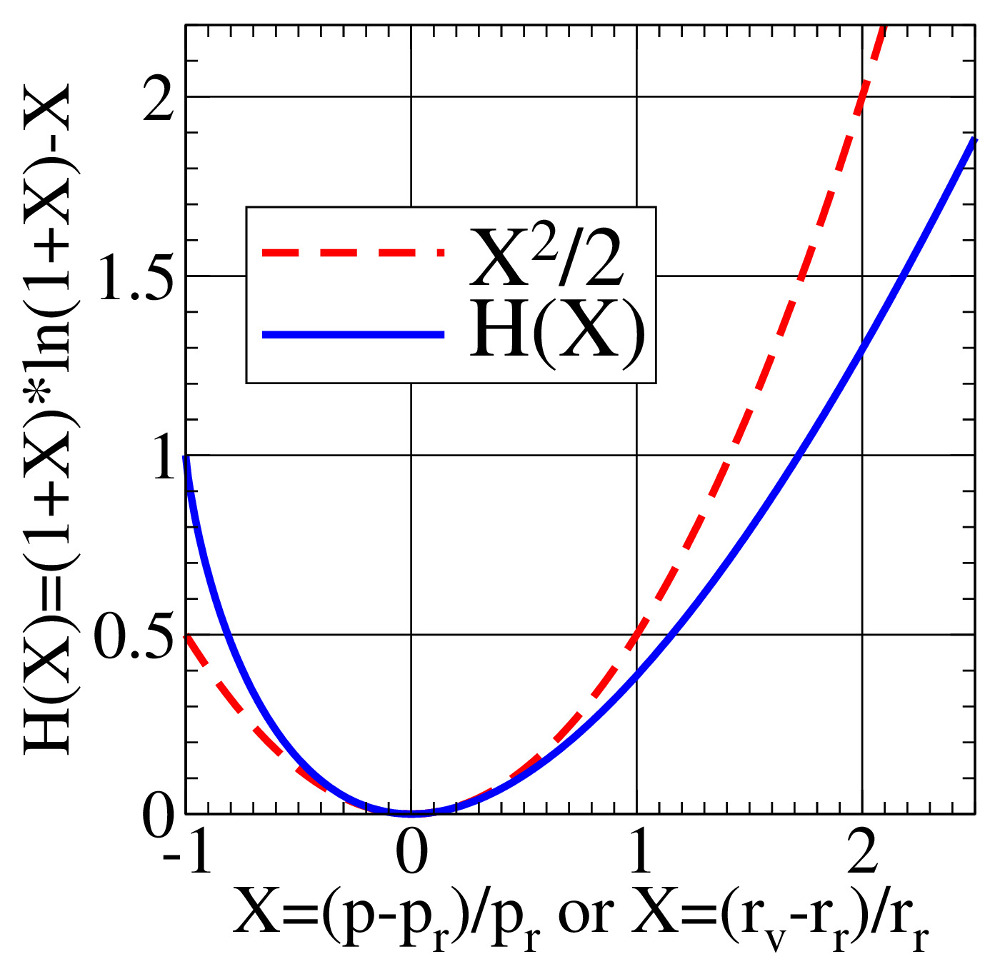}
\caption{\it \small 
The two functions $H(X)=(1\!+\!X)\ln(1\!+\!X)-X$ and $X^2/2$ plotted for $-1<X<+2.5$.
\label{figHX}}
\end{figure}

It is shown in Fig.~\ref{figHX} that $H(X)$ is positive and asymmetric with respect to $X=0$.
It is a quadratic-like function because $H(X)\approx X^2/2$ up to higher order terms.
$H(X)=0$ only for $X=0$, namely for $p=p_r$.

The constant reference pressure $p_r$ can enter the derivative in (\ref{def_ap_Xp}) and the term $p_r\:H(X_p)$ is equal to the function $p_r-p+p\:\ln(p\:/p_r) \approx (p-p_r)^2/(2\:p_r)$ called ``store of work for any layer under isothermal conditions'' in \citet{Margules_1901} and  studied in Eq.~(Ia)' page 505, the bottom of page 506 and the top of page 507 of this old paper.


\vspace{4mm}
\noindent
{\bf \underline{Appendix~E. The water components of $a_m$}.}
                             \label{appendixE}
\renewcommand{\theequation}{E-\arabic{equation}}
  \renewcommand{\thefigure}{E-\arabic{figure}}
   \renewcommand{\thetable}{E-\arabic{table}}
      \setcounter{equation}{0}
        \setcounter{figure}{0}
         \setcounter{table}{0}
\label{---Appendix-E}
\vspace{2mm}

The aim of this section is to show that $a_v$ given by (\ref{defav3}), which depends on the six pressures $p$, $p_r$, $p_d$,$(p_d)_r$, $e$ and $e_r$, can be expressed in terms of the sole water mixing ratios $r_v= q_v/q_d$ and $r_r$.
In this way, $a_v$ will be interpreted as the water-vapor component of $a_m$.

Division of (\ref{defaux2}) by (\ref{defaux1}) leads to  $e/p_d = r_v /r_0$, where $r_0 \equiv \: R_d/R_v = 0.622$.
The same result is valid for the reference state, leading to $e_r/(p_d)_r = r_r /r_0$ and to a  reference value for the mixing ratio given by
\vspace{-0.15cm}
\begin{equation}
   r_r \;  = \; r_0 \; e_r \:  / \, (p_d)_r  \: .
  \label{defrr}
\end{equation}
This reference mixing ratio is fully determined if $T_r$ and $p_r$ are known, because $p_r = (p_d)_r + e_r$ and $e_r(T_r)$ is the saturation pressure of water at $T_r$ and $(p_d)_r = p_r - e_r(T_r)$ can then be computed.

The four pressure terms involved in (\ref{defav3}) are computed by using $e/p_d = r_v /r_0$ and $e_r/(p_d)_r = r_r /r_0$, leading to
\vspace{-0.15cm}
\begin{align}
\!\!\!\!
  \frac{p_d}{p} \; &
              = \; \frac{p_d}{p_d + e} 
          \; = \;  \frac{r_0}{r_v + r_0} \, ,
 & \frac{p_r}{(p_d)_r} \; 
          \; = \;  \frac{r_r+r_0}{r_0} \, ,
     \label{def_am_aux1}
  \\
\!\!\!\!
  \frac{e}{p} \; &
              = \; \frac{e}{p_d + e} 
          \; = \;  \frac{r_v}{r_v + r_0} \, ,
 & \frac{p_r}{e_r} \; 
          \; = \;  \frac{r_r+r_0}{r_r} \, .
     \label{def_am_aux2}
\end{align}
The component $a_v$ given by  (\ref{defav3})  can then be written as
\vspace{-0.15cm}
\begin{align}
\!\!\!\!\!\!\!\!
  a_v & = R \: T_r \!
      \left[ \,
         \left(\frac{r_v}{r_v+r_0}\right)
         \ln\!\left(\frac{r_v}{r_{r}}\right)
    \! -  \ln\!\left(\frac{r_v+r_0}{r_{r}+r_0}\right)
      \, \right] \! .
  \!\!\!\!
   \label{def_am6}
\end{align}
This formulation of $a_v$ has already been derived in the exergetic analysis of moist-air processes described (in German) in \citet[Eq.~(10)]{Szargut_Styrylska_1969} and recalled in \citet[Eq.~(5.48), p.207]{Bejan_2016}, though with different notations.

The bracketed terms in (\ref{def_am6}) only depend on $r_v$ and on the two known reference values $T_r$ and $p_r$, since the reference mixing ratio is $r_r = r_0 \: e_r(T_r) / [\: p_r - e_r(T_r) \: ]$.
Therefore, $a_v$ will be called the water-vapor component of $a_m$.

The impacts of $q_l$ and $q_i$ are not neglected up to this point, because the condensed water contents impact the gas constant $R$, which depends on $q_v$ and $q_d = 1 - q_v - q_l - q_i$.
Conversely, the bracketed terms in (\ref{def_am6}), which generates the quadratic-like part of $a_v$, do not depend on $q_l$ or $q_i$.
These results could not be expected and are just imposed by the exact computations.

Let us introduce the water variables 
\vspace{-0.15cm}
\begin{equation}
Z_v  =  \frac{e}{p}  = \frac{r_v}{r_v+r_0} \: , \; \:
Z_r   =  \frac{e_r}{p_r}  = \frac{r_r}{r_r+r_0}
    \label{defXZvc} \: .
\end{equation}
which are computed with (\ref{def_am_aux2}).
The water component $a_v$ given by (\ref{def_am6}) can be transformed into the sum of the two terms depending on the function $H$, leading to
\vspace{-0.25cm}
\begin{align}
 a_v 
      & =   R \;   T_r \; Z_r \; 
        H \! \left( X_v \right) 
         +  R \;  T_r \:
         \left( 1-Z_r \right) \,
        H \! \left( Y_v \right )
         ,
\label{def_av5}
\end{align}
where 
\vspace{-0.15cm}
\begin{align}
 X_v \; & =  \:  \frac{Z_v}{Z_r} - 1 
   \; = \; \frac{r_0}{r_r} \:
          \left( 
          \frac{r_v - r_r}{r_v+r_0} 
          \right)
\; ,  \label{def_Xv} \\
 Y_v \; & =  \:  \frac{(1-Z_v)}{(1-Z_r)} - 1
  \; = \; - \:
          \left( 
          \frac{r_v - r_r}{r_v+r_0} 
          \right)
\: . \label{def_Yv}
\end{align}
The equality of (\ref{def_av5}) with (\ref{def_am6}) can be checked by using basic algebra.
This result has been obtained via a lengthy trial and error process, with the aim of introducing any of the quadratic-like functions $F$ or $H$ of the variable $(r_v-r_r)/r_r$.

\begin{table}
\caption{\it \small 
The ratio $|X_v/Y_v| = [\: p_r-e_r(T_r) \:]/e_r(T_r)$ computed for several reference temperatures $T_r$ and pressures $p_r$.
See the Table~\ref{Table_rr} for values of $e_r(T_r)$.
\vspace*{1mm}
\label{Table_Yv_over_Xv}}
\centering
\begin{tabular}{|c|c|c|c|}
\hline 
   $|X_v/Y_v|$   & $367.8$~hPa  & $\bf 800$~hPa & $1000$~hPa \\ 
\hline 
    $250$~K      & $438$      & $953$     & $1193$  \\ 
    $270$~K      & $77$       & $169$     & $212$   \\ 
    $273.15$~K   & $59$       & $130$     & $163$   \\
    $280$~K      & $36$       & $80$      & $100$   \\ 
    $300$~K      & $9.4$      & $22$      & $27$ \\
    $350$~K      & $-$        & $0.94$    & $1.4$ \\
\hline 
\end{tabular}
\end{table}

The ratio $|X_v/Y_v| = r_0 / r_r = [\: p_r-e_r(T_r) \:] / e_r(T_r) $ shown in Table~\ref{Table_Yv_over_Xv} is computed for the set of reference values $T_r$ and $p_r$ used in Tables~\ref{Table_Tr_Pr} and \ref{Table_rr}.
The ratio is larger than 20 for $T_r \leq 300$~K and $p_r = 800$ or $1000$~hPa. 
This result justifies the name ``large'' and ``small'' given to $X_v$ and $Y_v$, respectively.

The higher temperature $T_r=350$~K leads to small values of $|X_v/Y_v|$ which are close to unity, with an undefined (negative) ratio for $367.8$~hPa.
Values of $T_r > 300$~K are thus beyond the scope of the next definition for the water component of the exergy norm, where both $r_v$ and $r_r$ are much lower than  $r_0 \approx 622$~g~kg${}^{-1}$ only for $T_r \leq 300$~K, leading to $X_v \approx (r_v-r_r)/r_r$ and $Y_v \approx - \: (r_v-r_r)/r_0$.
The best candidate for a water dimensionless variable similar to $X_T = (T-T_r)/T_r$ is thus the large component $X_v$.


\vspace{4mm}
\noindent
{\bf \underline{App.~F. Separating properties of $F$ and $H$}.}
                           \label{appendixF}
\renewcommand{\theequation}{F-\arabic{equation}}
     \renewcommand{\thefigure}{F-\arabic{figure}}
       \renewcommand{\thetable}{F-\arabic{table}}
      \setcounter{equation}{0}
        \setcounter{figure}{0}
         \setcounter{table}{0}
\label{---Appendix-F}
\vspace{2mm}

Previous results cannot be used as such by replacing the terms $(T-T_r)^2$, $(p_s-p_r)^2$ and $(r_v-r_r)^2$ by the departure terms $(T')^2$, $(p'_s)^2$ and $(r'_v)^2$, respectively.
This issue is motivated by the usual applications where the perturbation terms $T'$, $p'_s$ and $r'_v$ may need to get zero average values, whereas $T-T_r$, $p_s-p_r$ and $r_v-r_r$ cannot cancel for all vertical levels and for constant values of $T_r$, $p_r$ and $r_r$.

It is thus important to introduce the mean values $\overline{T}$, $\overline{p_s}$ and $\overline{r_v}$ which denote averages of $T$, $p_s$ and $r_v$ computed for a given circle of latitudes, or for a given pressure level, or for any other kind of average like those considered in Fig.~F1.
The eddy departure terms will then be defined in the usual way by $T' = T - \overline{T}$, $p'_s = p_s - \overline{p_s}$ and $r'_v = r_v - \overline{r_v}$.

\begin{figure}[hbt]
\centering
\includegraphics[width=0.99\linewidth]{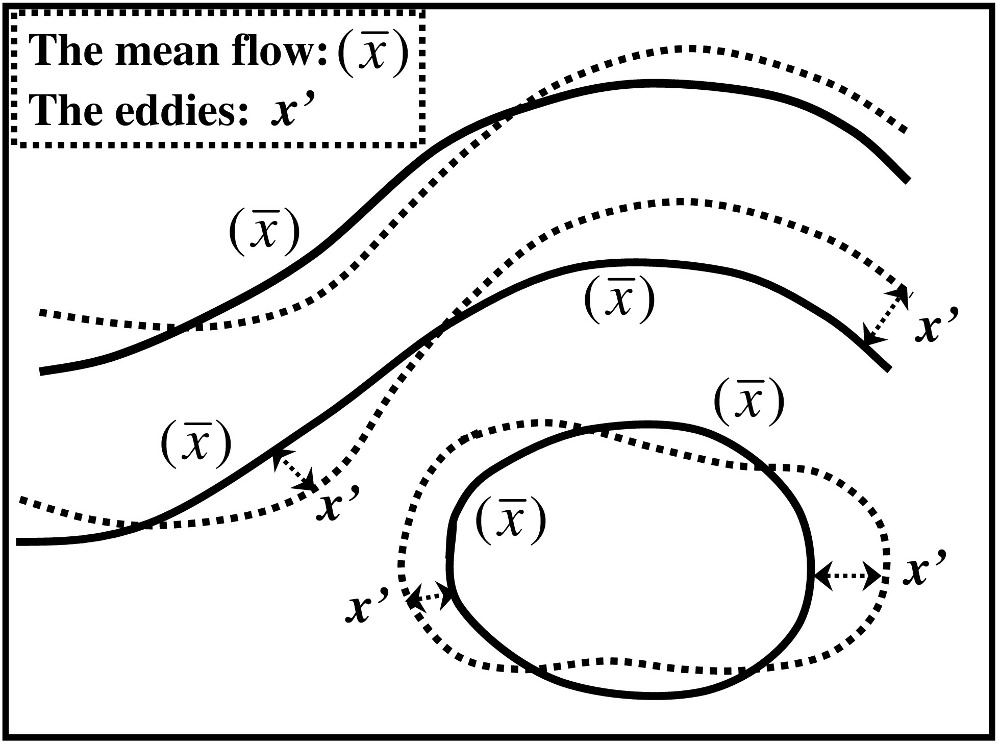}
\caption{\it \small 
The separation of the flow into an uneven basic state ($\overline{x}$, solid lines) plus the eddies (dashed lines), defined by $x' \equiv x  -  \overline{x}$. 
The $x$ term stands for the meteorological variables $T$, $p$, $Z_v$ or $r_v$, also $u$ and $v$.
\label{fig_Mean_Eddy}}
\end{figure}

Therefore, the aim is to express the available-enthalpy functions $a_T$, $a_p$ and $a_v$ depending on $T-T_r$, $p_s-p_r$ and $r_v-r_r$ in terms of the  ``energies of the mean state'' which depend on $(\overline{T}-T_r)^2/2$, $(\overline{p_s}-p_r)^2/2$ and $(\overline{r_v}-r_r)^2/2$ plus the ``energies of the eddies'' which depend on $\overline{(T')^2}/2$, $\overline{(p'_s)^2}/2$ and $\overline{(r'_v)^2}/2$.

For pure quadratic quantities, such as the kinetic energy, the basic separating property is given by the binomial law
\vspace{-0.15cm}
\begin{align}
 (X_1 + X_2)^2 & = (X_1)^2 + (X_2)^2 + 2\:X_1\:X_2 
  \: . \label{defX1X2a}
\end{align}
If the flow $X$  is separated into a mean part $X_1$ for which $\overline{X_1} \equiv X_1$, plus an eddy part $X_2$ for which $\overline{X_2} \equiv 0$, the separating property writes
\vspace{-0.15cm}
\begin{align}
 \overline{(X_1 + X_2)^2} & = \overline{(X_1)^2} +  \overline{(X_2)^2} 
  \: . \label{defX1X2b}
\end{align}

A similar exact separating property is derived for $F(X)$ in \citet{Marquet91,Marquet03}, and the one valid for $H(X)$ is shown in this Appendix.
For any variable written as $X = X_1 + X_2 + X_1 \: X_2$ the two properties
\vspace{-0.15cm}
\begin{align}
\! \! \!
 F(X)
     & =  F(X_1) 
        \: + \: F(X_2) 
        \: + \: X_1\:X_2 \: ,
  \label{defFX1X2a} \\
\! \! \!
 H(X) 
     & = (1+X_2)\:H(X_1)
    + (1+X_1)\:H(X_2)
    +  X_1 \: X_2 \: ,
  \label{defHX1X2a}
\end{align}
are valid for $X_1 > -1$ and $X_2 > -1$, which means $X_1 + X_2 + X_1\:X_2 = (1+X_1)(1+X_2)-1 \: > -1$. 
The flow $X$ is then separated into the same mean and eddy parts used to derived (\ref{defX1X2b}) and with $\overline{X_1} \equiv X_1$  and $\overline{X_2} \equiv 0$, leading to
\vspace{-0.15cm}
\begin{align}
 \overline{F(X)} \:
     & = \:     \overline{F(X_1)}
        \: + \: \overline{F(X_2)} 
   \: , \label{defFX1X2b} \\
 \overline{H(X)} \:
     & = \: \overline{H(X_1)}
    \: + \: (1+\overline{X_1})\;\;\overline{H(X_2)}
  \: . \label{defHX1X2b}
\end{align}
The physical consequence of (\ref{defX1X2b}),  (\ref{defFX1X2b}) and (\ref{defHX1X2b}) is the appearance of exact self-similarity properties verified by the total, mean and eddy parts of the flow: quadratic $F$ or $H$ functions generate quadratic $F$ or $H$ functions for the mean and the eddy parts of the flow.
More precisely, the quadratic approximation of (\ref{defFX1X2b}) will allow computations of $(T-T_r)^2/2$ in terms of $(\overline{T}-T_r)^2/2$ and $(T')^2/2$, with similar results derived from (\ref{defHX1X2b}) and valid for surface pressure and water-vapor mixing ratio.


\vspace{4mm}
\noindent
{\bf \underline{App.~G. Mean and eddy components of} \\
     \hspace*{16mm} \underline{$a_T$, $a_p$, $a_v$}.}
                           \label{appendixG}
\renewcommand{\theequation}{G-\arabic{equation}}
     \renewcommand{\thefigure}{G-\arabic{figure}}
       \renewcommand{\thetable}{G-\arabic{table}}
      \setcounter{equation}{0}
        \setcounter{figure}{0}
         \setcounter{table}{0}
\vspace{2mm}

Mean and eddy components of $a_T$ given by  (\ref{def_at_Xt}) can be computed by replacing $X_T  \:  = \: T/\,T_r - 1$  in (\ref{def_FX}) by
\vspace{-0.15cm}
\begin{align}
\!\!\! \!\!\!\!
  X_T  & = 
     \left(
     \frac{T'}{\;\overline{T}\;}
     \right)
     + 
     \left(
     \frac{\;\overline{T}\;}{T_r} - 1 
     \right)
     + 
     \left(
     \frac{T'}{\;\overline{T}\;}
     \right)
     \left(
     \frac{\;\overline{T}\;}{T_r} - 1 
     \right) \! ,
  \!\!
  \label{def_XT_mean_eddy}
\end{align}
where ${\overline{T}}/{T_r} - 1$ and  ${T}/\,{\overline{T}} - 1 = T' /\, {\overline{T}}$ correspond to $X_1$ and $ X_2$ in (\ref{defFX1X2b}), respectively.
It is then assumed that $c_p \approx  c_{pd}$ and $F(X)\approx X^2/2$, leading to
\vspace{-0.15cm}
\begin{align}
  \overline{a_T}  & \; \approx \;
        c_{pd} \; T_r \;
        F \!\left( \frac{\;\overline{T}\;}{T_r} - 1 \right)
        \: + \;
        c_{pd} \; T_r \;
       \overline{
        F \!\left( \frac{T'}{\;\overline{T}\;} \right)
        }, 
    \label{def_aT_mean} \\
  \overline{a_T}   & \; \approx \;
        c_{pd} \; 
              \frac{(\overline{T} - T_r)^2}{2 \: T_r}
        \: + \;
        c_{pd}
         \left(  
             \frac{T_r}{\; \overline{T} \;}
         \right)^{\! \! 2} \;
       \overline{
        \frac{(T')^2}{2 \: T_r}
        }  \; .
      \label{def_aT_eddy} 
\end{align}

The three dimensional integral of the first quadratic term in the r.h.s. of  (\ref{def_aT_eddy}) represents the ``unavailable enthalpy'' of the mean state $\overline{T}$ with respect to the isothermal reference state $T_r$.
The integral of the second quadratic term represents the ``available enthalpy'' of the perturbations $T'$ of the actual state $T$ with respect to the mean state $\overline{T}$, and it forms the temperature contribution of the squared norm which can be written as
\vspace{-1mm}
\begin{equation}
    N_T \;
       \equiv \;  \int\!\!\!\!\int\!\!\!\!\int \:
              \frac{c_{pd} \: T_r}{(\: \overline{T} \:)^2} \:
              \frac{(T')^2}{2} \:
              \frac{dm}{\Sigma}
      \label{def_N_T}  \: .
\end{equation}
This squared norm is studied in 
Sections~\ref{subsection_theory_new_norm} and
\ref{subsection_result_ARPEGE_tempe}.

The integral of $a_p$ given by (\ref{def_ap_Xp}) is computed by assuming that $R \approx R_d$, leading to
\vspace{-0.15cm}
\begin{align}
 A_p \; &  \approx \: R_d \: T_r \:  p_r
    \int\!\!\!\!\int \!  
    \left[ \:
      \int_{0}^{p_s}
      \frac{\partial \, H(X_p)}{\partial p } \:
      \frac{dp}{g} \:
    \right]
      \frac{d\Sigma}{\Sigma}
   \label{def_Np_approx_1} \: , \\
   A_p \; & \approx \: R_d \: T_r
    \int\!\!\!\!\int \!      
    \left[ \;
         H(X_{p_s}) - 1
    \; \right] 
    \: \frac{p_r}{g} \:
      \frac{d\Sigma}{\Sigma}
   \label{def_Np_approx_2} \: , 
\end{align}
where $X_{p_s} = {p_s}/{p_r} - 1$.
The term $-1$ is due to $X_p=-1$ for $p=0$ and $H(-1) = 1$, leading to a constant value $R_d \: T_r \: p_r / g$ which will not enter the definition of the squared norm component for pressure.
The aim is thus to compute mean and eddy components of
\vspace{-0.1cm}
\begin{align}
   B_p \: = \: 
   A_p \: + \: 
   R_d \; T_r \; \frac{p_r}{g} 
   &  \; \approx \; 
      R_d \; T_r
    \; \frac{p_r}{g} \; \;
    \overline{H(X_{p_s})}
   \label{def_Np_approx_3} \: , 
\end{align}
by replacing $X_{p_s}$ by
\vspace{-0.1cm}
\begin{align}
\!\!\! \!\!\!\!
  X_{p_s}  &\! = \!
     \left(
     \frac{p'_s}{\;\overline{p_s}\;}
     \right)
     +\!
     \left(
     \frac{\;\overline{p_s}\;}{p_r} - 1 
     \right)
     +\!
     \left(
     \frac{p'_s}{\;\overline{p_s}\;}
     \right)\!
     \left(
     \frac{\;\overline{p_s}\;}{p_r} - 1 
     \right) \! ,
\!\!\!\!
  \label{def_Xp_mean_eddy}
\end{align}
where ${\overline{p_s}}/{p_r} - 1$ and  ${p_s}/\,{\overline{p_s}} - 1 = p'_s /\, {\overline{p_s}}$ correspond to $X_1$ and $ X_2$ in (\ref{defHX1X2b}), respectively.
The separating property (\ref{defHX1X2b}) can then be applied to (\ref{def_Np_approx_3}), leading to
\vspace{-0.2cm}
\begin{align}
\!\!\! \!\!
 B_p  & \approx 
      R_d \: T_r \:
     \frac{p_r}{g}
  \left[ 
    H \! \left(\frac{\overline{p_s}}{p_r} - 1 \right)
    + 
    \left(
    \frac{\overline{p_s}}{p_r}
    \right) \:
    \overline{
      H \! \left(
           \frac{p'_s}{\overline{p_s}} 
      \right)}
  \: \right] \! ,
  \!\!\!
   \label{def_Np_approx_4} \\
\!\!\! \!\!
B_p  & \approx 
    \; \frac{R_d \; T_r}{g \; p_r} \;
    \frac{(\overline{p_s}-p_r)^2}{2}
    \: + \:
    \; \frac{R_d \; T_r}{g \; \overline{p_s}} \;
    \frac{\overline{(p'_s)^2}}{2}
   \label{def_Np_mean_eddy} ,
\end{align}
where it is assumed that $H(X)\approx X^2/2$.

The first quadratic term of $B_p$ in the r.h.s. of (\ref{def_Np_mean_eddy}) represents the unavailable enthalpy of the mean state $\overline{p_s}$ with respect to the constant reference pressure $p_r$.
The second quadratic term represents the available enthalpy of the perturbations $p'_s$ of the actual state $p_s$ with respect to the mean state $\overline{p_s}$.
This pressure contribution of the squared norm can be transformed back into a three dimensional integral, leading to
\vspace{-0.15cm}
\begin{align}
\!\!\! \!\!
   N_p 
   & \: \equiv \;
              \frac{R_d \, T_r}{g \: \overline{p_s}} \: 
              \frac{\overline{{(p'_s)}^2}}{2}
      = \! \int\!\!\!\!\int \limits^{ } \:
       \frac{R_d \; T_r}{(\overline{p_s})^2} \;
       \: \frac{p_s}{g} \;
       \frac{\overline{(p'_s)^2}}{2}
              \frac{d\Sigma}{\Sigma} ,
      \!\!\!
      \label{def_N_p}
      \\
\!\!\! \!\!
   N_p 
   & \: = \; \int\!\!\!\!\int \limits^{ } \:
       \frac{R_d \; T_r}{(\overline{p_s})^2} \;
     \left( \int_0^{p_s} \frac{dp}{g} \right) \;
       \frac{\overline{(p'_s)^2}}{2}
         \: \frac{d\Sigma}{\Sigma} ,
      \!\!\!
      \label{def_N_p2}
\end{align}
\vspace{-0.3cm}
\begin{equation}
   N_p 
   \; = \; \int\!\!\!\!\int\!\!\!\!\int\limits^{ } \:
       \frac{R_d \; T_r}{(\overline{p_s})^2} \;
       \frac{\overline{(p'_s)^2}}{2}
         \: \frac{dm}{\Sigma}
      \label{def_N_p3} \: . 
\end{equation}
This squared norm is studied in 
Section~\ref{subsection_theory_new_norm}.

It is shown in Appendix~E that the first term in the r.h.s. of (\ref{def_av5}) is much larger than the second term, due to $|X_v|\gg|Y_v|$.
This result is used together with the assumptions $R \approx R_d$, $r_0 \gg r_v$ and $r_0 \gg r_r$, leading to $Z_v \approx r_v / r_0$, $Z_r \approx r_r / r_0$ and $X_v \approx r_v / r_r - 1$, to approximate the (isobaric, horizontal or uneven) surface mean value of $a_v$ by 
\vspace{-1.mm}
\begin{align}
\overline{a_v} & \: \approx \: R_v \; T_r \; r_r \; \overline{H\left(\frac{r_v}{r_r} - 1\right)}
\: ,
\label{def_av_approx_3}
\end{align}
where $R_v = R_d / r_0$ has been used.

The separating property (\ref{defHX1X2b}) can then be applied to (\ref{def_av_approx_3}) and with the exact property
\vspace{-1mm}
\begin{align}
\!\!\! \!\!
     \left(
          \frac{r_v}{r_r} - 1 \!
     \right)
     & \! = 
     \left(
         \frac{r'_v}{\overline{r_v}}
     \right)
      \! + \! 
     \left(
     \frac{\overline{r_v}}{r_r} - 1  \!
     \right)
      \! + \!
     \left(
         \frac{r'_v}{\overline{r_v}}
     \right) \!
     \left(
     \frac{\overline{r_v}}{r_r} - 1 \!
     \right)\!\! ,
\!\!\!
  \label{def_Xv_mean_eddy}
\end{align}
where $r'_v = r_v - \overline{r_v}$.
The terms ${\overline{r_v}}/{r_r} - 1$ and  ${r_v}/\,{\overline{r_v}} - 1 = r'_v /\, {\overline{r_v}}$ correspond to $X_1$ and $ X_2$ in (\ref{defHX1X2b}), respectively, with the property $\overline{r'_v} = 0$ leading to
\vspace{-1mm}
\begin{align}
\!\!\! \!\!
 \overline{a_v}  & \approx 
      R_v \: T_r \: r_r 
  \left[
    H \! \left(\frac{\overline{r_v}}{r_r} - 1 \right)
     + 
    \left(
    \frac{\overline{r_v}}{r_r}
    \right) \;
    \overline{
      H \! \left(
           \frac{r'_v}{\overline{r_v}} 
      \right)}
  \: \right] \! .
\!\!
   \label{def_av_approx_4}
\end{align}
It is finally assumed that $H(X)\approx X^2/2$, leading to
\vspace{-1mm}
\begin{align}
 \overline{a_v} \; & \approx \; 
    \frac{R_v \; T_r}{r_r} \;
    \frac{(\overline{r_v}-r_r)^2}{2}
    \: + \:
    \; \frac{R_v \; T_r}{\overline{r_v}} \;
    \frac{\overline{(r'_v)^2}}{2}
   \label{def_av_approx_5} \: .
\end{align}

The integral of the first quadratic term in the r.h.s. of (\ref{def_av_approx_5}) represents the unavailable enthalpy of the mean state $\overline{r_v}$ with respect to the constant reference pressure $r_r$.
The integral of the second quadratic term represents the available enthalpy of the perturbations $r'_v$ of the actual state $r_v$ with respect to the mean state $\overline{r_v}$, and it forms the water contribution of the squared norm, which can be written as
\vspace{-1mm}
\begin{equation}
     N_v \;
       \equiv \;  \int\!\!\!\!\int\!\!\!\!\int \:
              \frac{R_v \: T_r}{\overline{r_v}} \:
              \frac{(r'_v)^2}{2} \:
              \frac{dm}{\Sigma} \: .
      \label{def_N_v}
\end{equation}    
This squared norm is studied in 
Sections~\ref{subsection_theory_new_norm},
\ref{subsection_result_ARPEGE_water},
\ref{subsection_result_CMC_GEOS} and
\ref{subsection_result_FSOI}.

If the exact moist value $R = (1-q_t)\:R_d + q_v \: R_v$ was not approximated by $R_d$ in (\ref{def_av5}), leading to $R_d/r_0 = R_v$ in (\ref{def_av_approx_3})-(\ref{def_av_approx_5}), then a factor $(1 + 2 \: \delta \: \overline{r_v})$ would exist (computations not shown) in the factor of $R_v$ in (\ref{def_N_v}), but leading to small terms in comparison with the definition (\ref{def_N_v}) for $N_v$.


\bibliographystyle{ametsoc2014}
\bibliography{Marquet_MWR-D-19-0081-R2}

\begin{thebibliography}{71}
\providecommand{\natexlab}[1]{#1}
\providecommand{\url}[1]{\texttt{#1}}
\renewcommand{\UrlFont}{\rmfamily}
\providecommand{\urlprefix}{URL }
\expandafter\ifx\csname urlstyle\endcsname\relax
  \providecommand{\doi}[1]{doi:\discretionary{}{}{}#1}\else
  \providecommand{\doi}{doi:\discretionary{}{}{}\begingroup
  \urlstyle{rm}\Url}\fi
\providecommand{\eprint}[2][]{\url{#2}}

\bibitem[{Baker and Daley(2000)Baker, and Daley}]{Baker_Daley00}
Baker, N.~L., and R.~Daley, 2000: Observation and background adjoint
  sensitivity in the adaptive observation targeting problem. \textit{Quart. J.
  Roy. Meteorol. Soc.}, \textbf{126~(565)}, 1431--1454,
  \doi{10.1002/qj.49712656511}.

\bibitem[{Barkmeijer et~al.(1998)Barkmeijer, Bouttier,, and
  Van~Gijzen}]{Barkmeijer_al98}
Barkmeijer, J., F.~Bouttier, and M.~Van~Gijzen, 1998: Singular vectors and
  estimates of the analysis-error covariance metric. \textit{Quart. J. Roy.
  Meteorol. Soc.}, \textbf{124~(549)}, 1695--1713,
  \doi{10.1002/qj.49712454916}.

\bibitem[{Barkmeijer et~al.(2001)Barkmeijer, Buizza, Palmer, Puri,, and
  Mahfouf}]{Barkmeijer_al01}
Barkmeijer, J., R.~Buizza, T.~N. Palmer, K.~Puri, and J.-F. Mahfouf, 2001:
  Tropical singular vectors computed with linearized diabatic physic.
  \textit{Quart. J. Roy. Meteorol. Soc.}, \textbf{127~(572)}, 685--708,
  \doi{10.1002/qj.49712757221}.

\bibitem[{Bejan(2016)}]{Bejan_2016}
Bejan, A., 2016: \textit{Advanced engineering thermodynamics}, 740 pp. {J}ohn
  {W}iley \& {S}ons, {I}nc.

\bibitem[{Borderies et~al.(2019)Borderies, Caumont, Delano\"e, Ducrocq,
  Fourri\'e,, and Marquet}]{Borderies_2019}
Borderies, M., O.~Caumont, J.~Delano\"e, V.~Ducrocq, N.~Fourri\'e, and
  P.~Marquet, 2019: Impact of airborne cloud radar reflectivity data
  assimilation on kilometre-scale numerical weather prediction analyses and
  forecasts of heavy precipitation events. \textit{Nat. Hazards Earth Syst.
  Sci.}, \textbf{19~(4)}, 907--926, \doi{10.5194/nhess-19-907-2019}.

\bibitem[{Buizza and Palmer(1995)Buizza, and Palmer}]{Buizza_Palmer_1995}
Buizza, R., and T.~N. Palmer, 1995: The singular-vector structure of the
  atmospheric global circulation. \textit{J. Atmos. Sci.}, \textbf{52~(9)},
  1434--1456, \doi{10.1175/1520-0469(1995)052<1434:TSVSOT>2.0.CO;2}.

\bibitem[{Buizza et~al.(1996)Buizza, Palmer, Barkmeijer, Gelaro,, and
  Mahfouf}]{Buizza_al_96}
Buizza, R., T.~N. Palmer, J.~Barkmeijer, R.~Gelaro, and J.-F. Mahfouf, 1996:
  Singular vector, norms and large-scale condensation. \textit{11th
  {C}onference on numerical weather prediction}, Norfolk, {V}irginia,
  {A}merican {M}eteorological {S}ociety, 50-52.

\bibitem[{Buizza et~al.(1993)Buizza, Tribbia, Molteni,, and
  Palmer}]{Buizza_al_93}
Buizza, R., J.~Tribbia, F.~Molteni, and T.~N. Palmer, 1993: Computation of
  optimal unstable structures for a numerical weather prediction model.
  \textit{Tellus A}, \textbf{45~(5)}, 388--407,
  \doi{10.1034/j.1600-0870.1993.t01-4-00005.x}.

\bibitem[{Cardinali(2009)}]{Cardinali_2009}
Cardinali, C., 2009: Monitoring the observation impact on the short-range
  forecast. \textit{Q. J. R. Meteorol. Soc.}, \textbf{135~(638)}, 239--250,
  \doi{10.1002/qj.366}.

\bibitem[{Chambon et~al.(2015)Chambon, Meunier, Guillaume, Piriou, Roca,, and
  Mahfouf}]{Chambon_al_2015}
Chambon, P., L.-F. Meunier, F.~Guillaume, J.-M. Piriou, R.~Roca, and J.-F.
  Mahfouf, 2015: Investigating the impact of the water-vapour sounding
  observations from saphir on board megha-tropiques for the arpege global
  model. \textit{Q. J. R. Meteorol. Soc.}, \textbf{141~(690)}, 1769--1779,
  \doi{10.1002/qj.2478}.

\bibitem[{C\^ot\'e et~al.(1998{\natexlab{a}})C\^ot\'e, Desmarais, Gravel,
  M\'ethot, Patoine, Roch,, and Staniforth}]{cote_al_1998b}
C\^ot\'e, J., J.-G. Desmarais, S.~Gravel, A.~M\'ethot, A.~Patoine, M.~Roch, and
  A.~Staniforth, 1998{\natexlab{a}}: The operational {CMC-MRB} global
  environmental multiscale ({GEM}) model. {P}art~{II}: {R}esults. \textit{Mon.
  Wea. Rev.}, \textbf{126~(6)}, 1397--1418,
  \doi{10.1175/1520-0493(1998)126<1397:TOCMGE>2.0.CO;2}.

\bibitem[{C\^ot\'e et~al.(1998{\natexlab{b}})C\^ot\'e, Gravel, M\'ethot,
  Patoine, Roch,, and Staniforth}]{cote_al_1998a}
C\^ot\'e, J., S.~Gravel, A.~M\'ethot, A.~Patoine, M.~Roch, and A.~Staniforth,
  1998{\natexlab{b}}: The operational {CMC-MRB} global environmental multiscale
  ({GEM}) model. {P}art~{I}: {D}esign considerations and formulation.
  \textit{Mon. Wea. Rev.}, \textbf{126~(6)}, 1373--1395,
  \doi{10.1175/1520-0493(1998)126<1373:TOCMGE>2.0.CO;2}.

\bibitem[{Courtier(1987)}]{Courtier_87}
Courtier, P., 1987: {\it Application du contr\^ole optimal \`a la pr\'evision
  num\'erique en M\'et\'eorologie (Application of the optimal control to the
  numerical forecast in meteorology)\/}. {PhD-thesis}, Paris-VI University,
  France.

\bibitem[{Courtier et~al.(1991)Courtier, Freyder, Geleyn, Rabier,, and
  Rochas}]{Courtier_al_1991}
Courtier, P., C.~Freyder, J.-F. Geleyn, F.~Rabier, and M.~Rochas, 1991: The
  arpege project at meteo france. \textit{Seminar on Numerical Methods in
  Atmospheric Models, 9-13 September 1991}, ECMWF, Shinfield Park, Reading,
  ECMWF, Vol.~II, 193-232, \urlprefix\url{https://www.ecmwf.int/node/8798}.

\bibitem[{Courtier et~al.(1994)Courtier, Th\'epaut,, and
  Hollingsworth}]{Courtier_al_1994}
Courtier, P., J.-N. Th\'epaut, and A.~Hollingsworth, 1994: A strategy for
  operational implementation of {4D-Var}, using an incremental approach.
  \textit{Quart. J. Roy. Meteorol. Soc.}, \textbf{120~(519)}, 1367--1387,
  \doi{10.1002/qj.49712051912}.

\bibitem[{Cover and Thomas(1991)Cover, and Thomas}]{Cover_Thomas_91}
Cover, T.~M., and J.~A. Thomas, 1991: \textit{Elements of Information Theory},
  563 pp. {J}ohn {W}iley \& {S}ons, {I}nc.

\bibitem[{Derber and Bouttier(1999)Derber, and Bouttier}]{Derber_Bouttier_99}
Derber, J., and F.~Bouttier, 1999: A reformulation of the background error
  covariance in the {ECMWF} global data assimilation system. \textit{Tellus A},
  \textbf{51~(2)}, 195--221, \doi{10.1034/j.1600-0870.1999.t01-2-00003.x}.

\bibitem[{Descamps et~al.(2007)Descamps, Ricard, Joly,, and
  Arbogast}]{Descamps_al_2007}
Descamps, L., D.~Ricard, A.~Joly, and P.~Arbogast, 2007: Is a real cyclogenesis
  case explained by generalized linear baroclinic instability? \textit{J.
  Atmos. Sci.}, \textbf{64~(12)}, 4287--4308, \doi{10.1175/2007JAS2292.1}.

\bibitem[{Ehrendorfer(2000)}]{Ehrendorfer_2000}
Ehrendorfer, M., 2000: The total energy norm in a quasigeostrophic model.
  \textit{J. Atmos. Sci.}, \textbf{57~(10)}, 3443--3451,
  \doi{10.1175/1520-0469(2000)057<3443:NACTEN>2.0.CO;2}.

\bibitem[{Ehrendorfer and Errico(1995)Ehrendorfer, and
  Errico}]{Ehrendorfer_Errico_1995}
Ehrendorfer, M., and R.~M. Errico, 1995: Mesoscale predictability and the
  spectrum of optimal perturbations. \textit{J. Atmos. Sci.}, \textbf{52~(20)},
  3475--3500, \doi{10.1175/1520-0469(1995)052<3475:MPATSO>2.0.CO;2}.

\bibitem[{Ehrendorfer et~al.(1999)Ehrendorfer, Errico,, and
  Reader}]{Ehrendorfer_al_1999}
Ehrendorfer, M., R.~M. Errico, and K.~D. Reader, 1999: Singular-vector
  perturbation growth in a primitive equation model with moist physics.
  \textit{J. Atmos. Sci.}, \textbf{56~(11)}, 1627--1648,
  \doi{10.1175/1520-0469(1999)056<1627:SVPGIA>2.0.CO;2}.

\bibitem[{Ehrendorfer and Tribbia(1997)Ehrendorfer, and
  Tribbia}]{Ehrendorfer_Tribbia_97}
Ehrendorfer, M., and J.~Tribbia, 1997: Optimal prediction of forecast error
  covariances through singular vectors. \textit{J. Atmos. Sci.},
  \textbf{54~(2)}, 286--313,
  \doi{10.1175/1520-0469(1997)054<0286:OPOFEC>2.0.CO;2}.

\bibitem[{Ehrendorfer et~al.(1995)Ehrendorfer, Tribbia,, and
  Errico}]{Ehrendorfer_al_95}
Ehrendorfer, M., J.~J. Tribbia, and R.~M. Errico, 1995: Mesoscale
  predictability: an assessment through adjoint methods. \textit{Seminar on
  predictability}, ECMWF, 157-183,
  \urlprefix\url{https://www.ecmwf.int/node/9270}.

\bibitem[{Eriksson and Lindgren(1987)Eriksson, and
  Lindgren}]{Eriksson_Lindgren_1987}
Eriksson, K.-E., and K.~Lindgren, 1987: Structural information in
  self-organizing systems. \textit{Physica Scripta}, \textbf{35~(3)}, 388--397,
  \doi{10.1088/0031-8949/35/3/026}.

\bibitem[{Eriksson et~al.(1987)Eriksson, Lindgren,, and
  M{\aa}nsson}]{Eriksson_al_1987}
Eriksson, K.-E., K.~Lindgren, and B.~{\AA}. M{\aa}nsson, 1987:
  \textit{Structure, Context, Complexity, Organization: physical aspects of
  information and value}, 446 pp. World Scientific Publishing Co. Pte. Ltd.,
  Singapore.

\bibitem[{Errico(2000)}]{Errico_2000}
Errico, R.~M., 2000: Interpretations of the total energy and rotational energy
  norms applied to determination of singular vectors. \textit{Q. J. R.
  Meteorol. Soc.}, \textbf{126~(566)}, 1581--1599,
  \doi{10.1002/qj.49712656602}.

\bibitem[{Errico and Ehrendorfer(1995)Errico, and
  Ehrendorfer}]{Errico_Ehrendorfer_95}
Errico, R.~M., and M.~Ehrendorfer, 1995: Moist singular vectors in a
  primitive-equation regional model. \textit{10th {C}onference on Atmospheric
  and oceanic waves and stability}, Big {S}ky, {MT}, {A}merican
  {M}eteorological {S}ociety, 235-238.

\bibitem[{Errico et~al.(2004)Errico, Reader,, and Ehrendorfer}]{Errico_al_2004}
Errico, R.~M., K.~D. Reader, and M.~Ehrendorfer, 2004: Singular vectors for
  moisture-measuring norms. \textit{Q. J. R. Meteorol. Soc.},
  \textbf{130~(598)}, 963--987, \doi{10.1256/qj.02.227}.

\bibitem[{Gauthier et~al.(2007)Gauthier, Tanguay, Laroche, Pellerin,, and
  Morneau}]{Gauthier_al_2007}
Gauthier, P., M.~Tanguay, S.~Laroche, S.~Pellerin, and J.~Morneau, 2007:
  Extension of {3DVAR} to {4DVAR}: Implementation of {4DVAR} at the
  {M}eteorological {S}ervice of {C}anada. \textit{Mon. Wea. Rev.},
  \textbf{135~(6)}, 2339--2354, \doi{10.1175/MWR3394.1}.

\bibitem[{Gelaro et~al.(2010)Gelaro, Langland, Pellerin,, and
  Todling}]{Gelaro_al_2010}
Gelaro, R., R.~H. Langland, S.~Pellerin, and R.~Todling, 2010: The {THORPEX}
  observation impact intercomparison experiment. \textit{Mon. Wea. Rev.},
  \textbf{138~(11)}, 4009--4025, \doi{10.1175/2010MWR3393.1}.

\bibitem[{Gelaro et~al.(2017)}]{Gelaro_al_2017}
Gelaro, R., and Coauthors, 2017: The modern-era retrospective analysis for
  research and applications, {V}ersion 2 ({MERRA-2}). \textit{J. Clim.},
  \textbf{30~(14)}, 5419--5454, \doi{10.1175/JCLI-D-16-0758.1}.

\bibitem[{Holdaway et~al.(2014)Holdaway, Errico, Gelaro,, and
  Kim}]{Holdaway_al_2014}
Holdaway, D., R.~Errico, R.~Gelaro, and J.~G. Kim, 2014: Inclusion of
  linearized moist physics in {NASA}'s {G}oddard earth observing system data
  assimilation tools. \textit{Mon. Wea. Rev.}, \textbf{142~(1)}, 414--433,
  \doi{10.1175/MWR-D-13-00193.1}.

\bibitem[{Holdaway et~al.(2015)Holdaway, Errico, Gelaro, Kim,, and
  Mahajan}]{Holdaway_al_2015}
Holdaway, D., R.~Errico, R.~Gelaro, J.~G. Kim, and R.~Mahajan, 2015: A
  linearized prognostic cloud scheme in {NASA}'s {G}oddard earth observing
  system data assimilation tools. \textit{Mon. Wea. Rev.}, \textbf{143~(10)},
  4198--4219, \doi{10.1175/MWR-D-15-0037.1}.

\bibitem[{Honerkamp(1998)}]{Honerkamp_1998}
Honerkamp, J., 1998: \textit{Statistical {P}hysics: an advanced approach with
  applications}, 410 pp. Springer-{V}erlag. {B}erlin {H}eildelberg
  {N}ew-{Y}ork.

\bibitem[{Janiskov{\'a} and Cardinali(2017)Janiskov{\'a}, and
  Cardinali}]{Janiskova_Cardinali_2017}
Janiskov{\'a}, M., and C.~Cardinali, 2017: On the impact of the diabatic
  component in the forecast sensitivity observation impact diagnostics (also:
  {ECMWF} technical memorandum {N}o.~786, 2016). \textit{{Data Assimilation for
  Atmospheric, Oceanic and Hydrologic Applications}}, Park, and Xu, Eds.,
  Vol.~III., Springer International Publishing, 483--511,
  \doi{10.1007/978-3-319-43415-5_22},
  \urlprefix\url{https://www.ecmwf.int/node/16716}.

\bibitem[{Joly(1995)}]{Joly_1995}
Joly, A., 1995: The stability of steady fronts and the adjoint method: Nonmodal
  frontal waves. \textit{J. Atmos. Sci.}, \textbf{52}, 3082--3108,
  \doi{10.1175/1520-0469(1995)052<3082:TSOSFA>2.0.CO;2}.

\bibitem[{Joly and Thorpe(1991)Joly, and Thorpe}]{Joly_Thorpe_1991}
Joly, A., and A.~J. Thorpe, 1991: The stability of time-dependent flows: An
  application to fronts in developing baroclinic waves. \textit{J. Atmos.
  Sci.}, \textbf{48}, 163--183,
  \doi{10.1175/1520-0469(1991)048<0163:TSOTDF>2.0.CO;2}.

\bibitem[{Karbou et~al.(2010)Karbou, G\'erard,, and Rabier}]{Karbou_al_2010}
Karbou, F., E.~G\'erard, and F.~Rabier, 2010: Global {4DVAR} assimilation and
  forecast experiments using {AMSU} observations over land. {P}art {I}: impacts
  of various land surface emissivity parameterizations. \textit{Wea.
  Forecasting}, \textbf{25}, 5--19, \doi{10.1175/2009WAF2222243.1}.

\bibitem[{Karlsson(1990)}]{Karlsson_90}
Karlsson, S., 1990: {\it Energy, Entropy and Exergy in the atmosphere\/}.
  {PhD-thesis}, Institute of Physical Resource Theory, Chalmers University of
  Technology. G\"oteborg, Sweden,
  \urlprefix\url{https://core.ac.uk/download/pdf/70599863.pdf}, 121 Pp.

\bibitem[{Kleeman(2002)}]{Kleeman_2002}
Kleeman, R., 2002: Measuring dynamical prediction utility using relative
  entropy. \textit{J.\ Atmos.\ Sci.}, \textbf{59~(13)}, 2057--2072,
  \doi{10.1175/1520-0469(2002)059<2057:MDPUUR>2.0.CO;2}.

\bibitem[{Kleist et~al.(2009)Kleist, Parrish, Derber, Treadon, Wu,, and
  Lord}]{Kleist_al_2009}
Kleist, D.~T., D.~F. Parrish, J.~C. Derber, R.~Treadon, W.-S. Wu, and S.~Lord,
  2009: Introduction of the {GSI} into the {NCEP} global data assimilation
  system. \textit{Wea. Forecasting}, \textbf{24~(6)}, 1691--1705,
  \doi{10.1175/2009WAF2222201.1}.

\bibitem[{Kullback(1959)}]{Kullback_1959}
Kullback, S., 1959: \textit{Information theory and statistics (1978, Dover
  Pub.)}, 409 pp. John Wiley \& Sons, Inc.

\bibitem[{Kullback and Leibler(1951)Kullback, and
  Leibler}]{Kullback_Leibler_1951}
Kullback, S., and R.~A. Leibler, 1951: On information and sufficiency.
  \textit{Ann. Math. Statist.}, \textbf{22}, 79--86,
  \doi{10.1214/aoms/1177729694}.

\bibitem[{Langland and Baker(2004)Langland, and Baker}]{Langland_Baker_2004}
Langland, R., and N.~Baker, 2004: Estimation of observation impact using the
  {NRL} atmospheric variational data assimilation adjoint system.
  \textit{Tellus~A}, \textbf{56}, \doi{10.3402/tellusa.v56i3.14413}.

\bibitem[{Lorenz(1955)}]{Lorenz_1955}
Lorenz, E.~N., 1955: Available potential energy and the maintenance of the
  general circulation. \textit{Tellus}, \textbf{7},
  \doi{10.3402/tellusa.v7i2.8796}.

\bibitem[{Lorenz(1978)}]{Lorenz_1978}
Lorenz, E.~N., 1978: Available energy and the maintenance of a moist
  circulation. \textit{Tellus}, \textbf{30}, \doi{10.3402/tellusa.v30i1.10308}.

\bibitem[{Lorenz(1979)}]{Lorenz_1979}
Lorenz, E.~N., 1979: Numerical evaluation of moist available energy.
  \textit{Tellus}, \textbf{31}, \doi{10.3402/tellusa.v31i3.10429}.

\bibitem[{Mahfouf and Bilodeau(2007)Mahfouf, and
  Bilodeau}]{Mahfouf_Bilodeau_2007}
Mahfouf, J.-F., and B.~Bilodeau, 2007: Adjoint sensitivity of surface
  precipitation to initial conditions. \textit{Mon. Wea. Rev.}, \textbf{135},
  \doi{10.1175/MWR3439.1}.

\bibitem[{Mahfouf and Buizza(1996)Mahfouf, and Buizza}]{Mahfouf_Buizza_96}
Mahfouf, J.-F., and R.~Buizza, 1996: On the inclusion of physical processes in
  linear forward and adjoint models. {I}mpact of large-scale condensation on
  singular vectors. \textit{Newsletter Number 72}, ECMWF, 2-6,
  \urlprefix\url{https://www.ecmwf.int/node/14652}.

\bibitem[{Mahfouf et~al.(1996)Mahfouf, Buizza,, and Errico}]{Mahfouf_al_96}
Mahfouf, J.-F., R.~Buizza, and R.~M. Errico, 1996: Strategy for including
  physical processes in the {ECMWF} variational data assimilation system.
  \textit{Workshop on non-linear aspects of data assimilation}, ECMWF, 595-632,
  \urlprefix\url{https://www.ecmwf.int/node/10924}.

\bibitem[{Majda et~al.(2002)Majda, Kleeman,, and Cai}]{Majda_al_2002}
Majda, A.~J., R.~Kleeman, and D.~Cai, 2002: A mathematical framework for
  quantifying predictability through relative entropy. \textit{Methods Appl.
  Anal.}, \textbf{9~(3)}, 425--444.

\bibitem[{Margules(1910)}]{Margules_1901}
Margules, M., 1910: The mechanical equivalent of any given distribution of
  atmospheric pressure, and the maintenance of a given difference in pressure
  ({T}ranslation by {C}. {A}bbe of a lecture read at the meeting of the
  imperial academy of science, {V}ienna, {J}uly, 11, 1901). \textit{Smithsonian
  Miscellaneous collections.}, \textbf{51~(4)}, 501--532,
  \urlprefix\url{https://www3.nd.edu/~powers/ame.20231/gibbs1873b.pdf}.

\bibitem[{Marquet(1991)}]{Marquet91}
Marquet, P., 1991: On the concept of exergy and available enthalpy: Application
  to atmospheric energetics. \textit{Quart. J. Roy. Meteorol. Soc.},
  \textbf{117~(499)}, 449--475, \doi{10.1002/qj.49711951112},
  \urlprefix\url{https://arxiv.org/abs/1402.4610}.

\bibitem[{Marquet(1993)}]{Marquet93}
Marquet, P., 1993: Exergy in meteorology: definition and properties of moist
  available enthalpy. \textit{Quart. J. Roy. Meteorol. Soc.},
  \textbf{119~(511)}, 567--590, \doi{10.1002/qj.49711951112},
  \urlprefix\url{https://arxiv.org/abs/1807.05830}.

\bibitem[{Marquet(2003)}]{Marquet03}
Marquet, P., 2003: The available-enthalpy cycle. {I}: {I}ntroduction and basic
  equations. \textit{Quart. J. Roy. Meteorol. Soc.}, \textbf{129~(593)},
  2445--2466, \doi{10.1256/qj.01.62},
  \urlprefix\url{https://arxiv.org/abs/1403.5671}.

\bibitem[{Marquet and Dauhut(2018)Marquet, and Dauhut}]{Marquet_Thibaut_2018}
Marquet, P., and T.~Dauhut, 2018: Reply to ``comments on 'a third-law
  isentropic analysis of a simulated hurricane'''. \textit{J. Atmos. Sci.},
  \textbf{75~(10)}, 3735--3747, \doi{10.1175/JAS-D-18-0126.1},
  \urlprefix\url{https://arxiv.org/abs/1805.00834}.

\bibitem[{Palmer et~al.(1998)Palmer, Gelaro, Barkmeijer,, and
  Buizza}]{Palmer_al_98}
Palmer, T.~N., R.~Gelaro, J.~Barkmeijer, and R.~Buizza, 1998: Singular-vectors,
  metrics, and adaptative observations. \textit{J. Atmos. Sci.},
  \textbf{55~(4)}, 633--653,
  \doi{doi:10.1175/1520-0469(1998)055<0633:SVMAAO>2.0.CO;2}.

\bibitem[{Pearce(1978)}]{Pearce_1978}
Pearce, R.~P., 1978: On the concept of available potential energy. \textit{Q.
  J. R. Meteorol. Soc.}, \textbf{104~(441)}, 737--755,
  \doi{10.1002/qj.49710444115}.

\bibitem[{Procaccia and Levine(1976)Procaccia, and
  Levine}]{Procaccia_Levine_1976}
Procaccia, I., and R.~D. Levine, 1976: Potential work: A statistical-mechanical
  approach for systems in disequilibrium. \textit{J. Chem. Phys.},
  \textbf{65~(8)}, 3357--3364, \doi{10.1063/1.433482}.

\bibitem[{Putman(2007)}]{Putman_2007}
Putman, W., 2007: {\it Development of the finite-volume dynamical core on the
  cubed-sphere\/}. {PhD-thesis}, Florida State University, U.S.A.,
  \urlprefix\url{https://diginole.lib.fsu.edu/islandora/object/fsu%3A168667},
  91 Pp.

\bibitem[{Rabier et~al.(1996)Rabier, Klinker, Courtier,, and
  Hollingsworth}]{Rabier_al_1996}
Rabier, F., E.~Klinker, P.~Courtier, and A.~Hollingsworth, 1996: Sensitivity of
  forecast errors to initial conditions. \textit{Q. J. R. Meteorol. Soc.},
  \textbf{122~(529)}, 121--150, \doi{10.1002/qj.49712252906}.

\bibitem[{Rivi\`ere et~al.(2009)Rivi\`ere, Lapeyre,, and
  Talagrand}]{Riviere_al_2009}
Rivi\`ere, O., G.~Lapeyre, and O.~Talagrand, 2009: A novel technique for
  nonlinear sensitivity analysis: Application to moist predictability.
  \textit{Q. J. R. Meteorol. Soc.}, \textbf{135~(643)}, 1520--1537,
  \doi{10.1002/qj.460}.

\bibitem[{Shannon(1948)}]{Shannon_1948}
Shannon, C.~E., 1948: A mathematical theory of communication. \textit{Bell
  System Technical Journal}, \textbf{27~(3)}, 379--423,
  \doi{10.1002/j.1538-7305.1948.tb01338.x}.

\bibitem[{Szargut and Styrylska(1969)Szargut, and
  Styrylska}]{Szargut_Styrylska_1969}
Szargut, J., and T.~Styrylska, 1969: Die exergetische {A}nalyse von {P}rozessen
  der feuchten {L}uft ({A}n exergetic analysis of processes for damp air).
  \textit{Heiz.-L\"uft.-Haustechn.}, \textbf{20~(5)}, 173--178.

\bibitem[{Talagrand(1981)}]{Talagrand_1981}
Talagrand, O., 1981: A study of the dynamics of four-dimensional data
  assimilation. \textit{Tellus.}, \textbf{33~(1)}, 43--60,
  \doi{10.3402/tellusa.v33i1.10693}.

\bibitem[{Th\'epaut and Courtier(1991)Th\'epaut, and
  Courtier}]{Thepaut_Courtier_1991}
Th\'epaut, J.-N., and P.~Courtier, 1991: Four-dimensional variational data
  assimilation using the adjoint of a multilevel promitive-equation model.
  \textit{Q. J. R. Meteorol. Soc.}, \textbf{117~(502)}, 1225--1254,
  \doi{10.1002/qj.49711750206}.

\bibitem[{Thomson(1853)}]{Thomson_1853}
Thomson, W., 1853: On the restoration of mechanical energy from an unequally
  heated space. \textit{Phil. Mag.}, \textbf{5~(30, 4e series)}, 102--105.

\bibitem[{Tr\'emolet(2008)}]{Tremolet_2008}
Tr\'emolet, Y., 2008: Computation of observation sensitivity and observation
  impact in incremental variational data assimilation. \textit{Tellus~A},
  \textbf{60~(5)}, 964--978, \doi{10.1111/j.1600-0870.2008.00349.x}.

\bibitem[{Wu et~al.(2002)Wu, Purser,, and Parrish}]{Wu_al_2002}
Wu, W.-S., R.~J. Purser, and D.~F. Parrish, 2002: Three-dimensional variational
  analysis with spatially inhomogeneous covariances. \textit{Mon. Wea. Rev.},
  \textbf{130~(12)}, 2905--2916,
  \doi{10.1175/1520-0493(2002)130<2905:TDVAWS>2.0.CO;2}.

\bibitem[{Xu(2006)}]{Xu_2006}
Xu, Q., 2006: Measuring information content from observations for data
  assimilation: relative entropy versus {S}hannon entropy difference.
  \textit{Tellus A}, \textbf{59~(2)}, 198--209,
  \doi{10.1111/j.1600-0870.2006.00222.x}.

\bibitem[{Zadra et~al.(2004)Zadra, Buenner, Laroche,, and
  Mahfouf}]{Zadra_al_2004}
Zadra, A., M.~Buenner, S.~Laroche, and J.-F. Mahfouf, 2004: Impact of the {GEM}
  model simplified physics on extratropical singular vectors. \textit{Q. J. R.
  Meteorol. Soc.}, \textbf{130~(602)}, 2541--2569, \doi{10.1256/qj.03.208}.

\end{thebibliography}

\end{document}